\documentclass[final,3p,times]{elsarticle}
\usepackage{float}
\usepackage{graphicx}
\usepackage[colorlinks,citecolor=blue,urlcolor=blue]{hyperref}
\usepackage{amssymb,amsmath}
\usepackage{color}
\usepackage{threeparttable}

\journal{\url{arXiv.org}}

\begin{document}

\begin{frontmatter}

\title{Respondent-driven Sampling on Directed Networks}

\author[kiphs,susocial]{Xin Lu\corref{cor1}}
\ead{lu.xin@sociology.su.se}
\author[sumath]{Jens Malmros}
\author[susocial]{Fredrik Liljeros}
\author[sumath]{Tom Britton}


\cortext[cor1]{Address for correspondence: Xin Lu, Department of Sociology, Stockholm University, SE-106 91, Stockholm, Sweden.}
\address[kiphs]{Department of Public Health Sciences, Karolinska Institute, Stockholm, Sweden}
\address[susocial]{Department of Sociology, Stockholm University, Stockholm, Sweden}
\address[sumath]{Department of Mathematics, Stockholm University, Stockholm, Sweden}

\begin{abstract}
Respondent-driven sampling (RDS) is a commonly used substitute for random sampling when studying hidden populations, such as injecting drug users or men who have sex with men, for which no sampling frame is known. The method is an extension of the snowball sample method and can, given that some assumptions are met, generate unbiased population estimates. One key assumption, not likely to be met, is that the acquaintance network in which the recruitment process takes place is undirected, meaning that all recruiters should have the potential to be recruited by the person they recruit. Here we investigate the potential bias of directedness by simulating RDS on real and artificial network structures. We show that directedness is likely to generate bias that cannot be compensated for unless the sampled individuals know how many that potentially may have recruited them (i.e. their indegree), which is unlikely in most situations. Based on one known parameter, we propose an estimator for RDS on directed networks when only outdegrees are observed.

By comparison of current RDS estimators' performances on networks with varying structures, we find that our new estimator, together with a recent estimator, which requires the population size as a known quantity, have relatively low level of estimate error and bias. Based on our new estimator, sensitivity analysis can be made by varying values of the known parameter to take uncertainty of network directedness and error in reporting degrees into account. Finally, we have developed a bootstrap procedure for the new estimator to construct confidence intervals.
\end{abstract}

\begin{keyword}
Respondent-driven Sampling\sep directed networks\sep HIV
\end{keyword}

\end{frontmatter}

\section{Introduction}
Hidden populations (hard-to-reach populations), such as injecting drug users (IDU), men who have sex with men (MSM), and sex workers (SW) and their sexual partners, are generally considered as critical actors in the HIV epidemic~\cite{UNAIDS2010, Malekinejad2008, Goel2009}. Consequently, obtaining population characteristics and risk behaviors of these populations are critical for developing efficient disease control strategies. However, the lack of sampling frames for such populations makes traditional estimation methods based on random samples practically useless. Other methods have been proposed for such situations, for example key informant sampling~\cite{Deaux1985}, targeted/location sampling~\cite{Watters1989} and snowball sampling~\cite{Erickson1979}.

A more recent method is Respondent Driven Sampling (RDS), which was proposed to overcome difficulties when sampling hidden populations~\cite{Heckathorn1997,Heckathorn2002,Volz2008}. The RDS method starts with an initial selection of respondents, which are called ``seeds''. Each seed is given a number of ``coupons'' -- tickets for participation in the study -- to distribute to friends and acquaintances within the population of interest. When interviewed (anonymously), a new respondent is in turn given coupons to distribute. Everyone is rewarded both for completing the interview, and for recruiting their peers into the study. If the recruitment chains are sufficiently long, the sample composition will stabilize and become independent of the seeds. Additionally, information about who recruits whom and each respondent's personal network size (degree) are recorded.

There are two significant improvements in RDS compared to other non-random methods that has been used when sampling hidden population: Firstly, it uses dual incentives to encourage respondents to recruit more peers into the study. Secondly, when several assumptions are fulfilled, unbiased estimates of population proportions can be obtained by utilizing the recruitment and degree information collected in the sample.

Suppose a RDS study is performed on a connected undirected network with the additional assumptions that:
\begin{description}
 \item [(i)] sampling of peer recruitment is done with replacement;
 \item [(ii)] each participant recruits one new participant to the study; and
 \item [(iii)] participants recruit randomly from their neighbors.
\end{description}
Then, the sampling probability of an individual $i$ will be proportional to its degree when the sample reaches equilibrium. The population fraction $p_A$ having a certain property $A$ (e.g.\ $p_A$ could denote the fraction among intravenous drug-users that are HIV-positive) can then be estimated by the weighted proportion of the sample fraction as in Volz and Heckathorn~\cite{Volz2008}:
\begin{equation}
{{\hat p_{A}^{VH_{out}}} = \frac{{\sum\limits_{{i} \in U \cap A} {{d_i}^{ - 1}} }}{{\sum\limits_{{i} \in U} {{d_i}^{ - 1}} }}},\label{eq_RDSII}
\end{equation}
where the sample population $U$ has been divided into two disjoint subsets $A$ and $B=A^C$ depending on the reported properties of respondents, and $d_i$ denotes the degree of individual $i$ in the sample.

Based on equating the number of crossrelations between subgroups of property $A$ and $B$, Salganik and Heckathorn~\cite{Salganik2004} proposed an another widely used estimator for $p_A$:
\begin{equation}
{\hat p_{A}^{SH_{out}}} = \frac{{{\hat{s}_{BA}}{{\hat{\bar D}}_B}}}{{{\hat{s}_{AB}}{{\hat{\bar D}}_A} + {\hat{s}_{BA}}{{\hat{ \bar D}}_B}}},\label{eq_RDSI}
\end{equation}
where ${\hat{\bar D}_A} = \frac{{{n_A}}}{{\sum\limits_{{i} \in U \cap A} {{d_i}^{ - 1}} }}$ and ${\hat{\bar D}_B} = \frac{{{n_B}}}{{\sum\limits_{{i} \in U \cap B} {{d_i}^{ - 1}} }}$ are the estimated harmonic mean degrees for the two subgroups. ($n_A$ and $n_B$ denote the number of $A$- and $B$-individuals in the sample respectively), and $\hat{s}_{BA}$ denotes the sample fraction of all $B$-respondents naming $A$-peers and similarly $\hat{s}_{AB}$ denotes the sample fraction of all $A$-respondents naming $B$-peers. For simplicity, we henceforth refer to \eqref{eq_RDSII} and \eqref{eq_RDSI} the $VH_{out}$ and $SH_{out}$ estimators respectively; the subscript \emph{out} indicates that respondents out-degrees have been recorded, which will be important when we move to directed networks later on.

The ability to produce unbiased population estimates and a feasible field implementation have contributed to a rapid increase in RDS studies conducted globally in recent years~\cite{Malekinejad2008, Johnston2008}. There has also been an increase in studies evaluating the performance of RDS estimators as well as in developing new estimators~\cite{Goel2010,Gile2010,Amber2010,Gile2011}.

Previous studies are mostly based on the assumption that relationships are reciprocal and the probability of being recruited by a peer is likewise equal on both ends of a relationship. Consequently, the network among which recruitments could take place is undirected. However, it is well-known that social networks, such as friendship networks, are generally directed to various extents. For example, in the study of Scott and Dana~\cite{South2004}, only 6,669 out of 12,931 ``best friend'' nominations were found to be reciprocal, and in the study conducted by Wallace~\cite{Wallace1966,Feld2002}, an average of 55.0 reciprocal nominations per respondent were found while the mean degree was 94.8. Evidence of irreciprocal recruiter-recruitee relationships has also been found in many RDS studies, e.g., in a RDS study of IDUs in Sydney, Australia~\cite{Dana2011}, 29\% of the respondents consider the relationship to their recruiter to be ``not very close'', and in a study of IDUs in Tijuana, Mexico~\cite{Daniela2009}, only 62\% of the respondents consider their relationship with their recruiter as "friend". Additionally, in a study for MSM in Beijing, China~\cite{Ma2007}, 8.5\% participants said they received their coupons from a stranger, and  between  3\% to 7\% of recruitments were found to be from strangers in the RDS studies on drug users and MSMs in three US cities and in St. Petersburg, Russia~\cite{Iguchi2009}.

In~\cite{Xin2011}, it has been shown that current RDS estimators may generate relatively large biases and errors if the studied networks are directed, indicating that estimates from previous RDS studies should be interpreted and generalized with caution. This study aims to further evaluate the influence of structural network properties on the performance of RDS estimators under the assumption that the underlying social network is (partially) directed, and to derive new estimators allowing networks to be directed. In our evaluation, we use simulated data, as well as a real online MSM social network, to generate networks with varying directedness, degree correlation, indegree-outdegree correlation and homophily. Based on one known parameter, we propose a new estimator for RDS on directed networks and show how sensitivity analysis can be used to generate estimate intervals induced by intervals of unobserved network properties. Additionally, we develop a bootstrap procedure for the new estimator to construct confidence intervals.

\section{RDS estimation on directed networks}\label{sec_rdsdirected}

We now investigate the properties of the RDS process on a directed network. For the purpose of this study, we focus on the problem of estimating the community fraction $p_A$ having a certain dichotomous property $A$. Let $G$ denote a (partially) directed network and let ${e_{ij}} = 1$ if there is a directed edge from $i$ to $j$ and ${e_{ij}} =0$ otherwise. A reciprocal edge between $i$ and $j$ is hence reflected by $e_{ij}=e_{ji}=1$. We assume that $G$ is strongly connected, i.e., there is a directed path between any pair of nodes -- otherwise we of course would not be able to to estimate $p_A$ well since it may then be impossible to reach certain parts of the community with RDS. Finally, we let $N$ denote the community size, most often an unknown quantity in hidden or hard-to-reach populations. In what follows, assumptions \textbf{(i)-(iii)} are assumed to be fulfilled in the RDS process.

\subsection{Extension of $VH_{out}$ estimator to directed networks}

When a RDS process takes place on a strongly connected network $G$, the recruitment of new respondents are dependent only on the current respondent, since he will select a new respondent uniformly from his peers. Thus, RDS possesses the Markov property~\cite{HASTINGS01041970} and can be modeled as a Markov process with transition matrix $R = \{ {a_{ij}} = {e_{ij}}/d_i^{out},1 \leqslant i,j \leqslant N\} $, where $d_i^{out}$ is the outdegree of node $i$ ~\cite{Volz2008,Xin2011}. This process has a unique equilibrium distribution $\pi  = [{\pi _1} \cdots {\pi _N}]$ satisfying ${R^T}\pi^T  = \pi^T $, indicating that $\pi $ is the eigenvector corresponding to eigenvalue 1 for ${R^T}$. Consequently, $\pi_i$ can be used to obtain the Hansen-Hurwitz estimator where observations are weighted by the inverse of the sampling probability~\cite{Xin2011}:

\begin{equation}{\hat p_{A}^{Eig}} = \frac{{\sum\limits_{{i} \in U \cap A} {{\pi _i}^{ - 1}} }}{{\sum\limits_{{j} \in U} {{\pi _j}^{ - 1}} }}\label{eq_RDSeig}.\end{equation}

It has been shown that when the network is undirected, $\pi_i  = {d_i}/\sum\limits_{j = 1}^N {{d_j}} $ is the analytical solution for $\pi$ and $\hat p_{A}$ can be estimated by the ${VH_{out}}$ estimator: $\hat p_A^{V{H_{out}}} = {{\sum\limits_{{i} \in U \cap A} {{d_i}^{ - 1}} } \mathord{\left/
 {\vphantom {{\sum\limits_{{i} \in U \cap A} {{d_i}^{ - 1}} } {\sum\limits_{{i} \in U} {{d_i}^{ - 1}} }}} \right.
 \kern-\nulldelimiterspace} {\sum\limits_{{i} \in U} {{d_i}^{ - 1}} }}$.

Unfortunately, no analytical solution for $\pi$ is available for a general directed network. However, note that under the above assumptions, the RDS process is merely a random walk on the network, for which we can easily adopt the mean field approach in~\cite{Fortunato2007} to derive an approximation of $\pi$:

Let ${\rm K} \equiv ({{\rm K}_{in}},{{\rm K}_{out}})$ be the set of nodes in the network with indegree ${{\rm K}_{in}}$ and outdegree ${{\rm K}_{out}}$, and let ${f_{\rm K}}$ be the proportion of ${\rm K}$-nodes in the set; then, the average inclusion probability of nodes in ${\rm K}$ is
\begin{equation}{\bar \pi ({\rm K}) \equiv \frac{1}{{N{f_{\rm K}}}}\sum\limits_{i \in {\rm K}} {{\pi _i}}}. \end{equation}

Using that we have a random walk assumed to be in equilibrium, and taking the average over all nodes of degree ${\rm K}$, we get
\begin{equation}{\bar \pi ({\rm K}) = \frac{1}{{N{f_{\rm K}}}}\sum\limits_{i \in {\rm K}}\sum\limits_{j:{\rm k}_{out}(j)\neq 0} {\frac{e_{ji}}{{\rm k}_{out}(j)}\pi_j}}, \end{equation}
where ${\rm k}_{out}(j)$ is the outdegree of node $j$.

Then, the sum over $j$ is parted into two, one over the degree classes ${{\rm K}'}$ and the other over the nodes within each degree class ${{\rm K}'}$. We substitute $\pi_j$ with the mean value within its degree class ${\rm K}'$, yielding

\begin{equation}{\bar \pi ({\rm K}) \simeq \frac{1}{{N{f_{\rm K}}}}\sum\limits_{{\rm K}'} {\frac{{\bar \pi ({\rm K}')}}{{{{{\rm K}'}_{out}}}}\sum\limits_{i \in {\rm K}} {\sum\limits_{j \in {\rm K}'} {{e_{ji}}} } } {\rm{    }}\\
{\rm{         }} = \frac{1}{{N{f_{\rm K}}}}\sum\limits_{{\rm K}'} {\frac{{\bar \pi ({\rm K}')}}{{{{{\rm K}'}_{out}}}}{E_{{\rm K}' \to {\rm K}}}},} \end{equation}
where ${{E_{{\rm K}' \to {\rm K}}}}$ is the total number of edges pointing from nodes of degree ${\rm K}'$ to nodes of degree ${\rm K}$, which we can write as ${E_{{\rm K}' \to {\rm K}}} = {{\rm K}_{in}}f_{\rm K}N\frac{{{E_{{\rm K}' \to {\rm K}}}}}{{{{\rm K}_{in}}f_{\rm K}N}}= {{\rm K}_{in}}f_{\rm K}Nf_{{\rm K}'|{\rm K}}$, where $f_{{\rm K}'|{\rm K}}$ is the proportion of edges pointing to nodes in ${\rm K}$ originating in ${\rm K}'$.

We finally obtain
\begin{equation}{\bar \pi ({\rm K}) = {{\rm K}_{in}}\sum\limits_{{\rm K}'} {\frac{{f_{{\rm K}'|{\rm K}}}}{{{{{\rm K}'}_{out}}}}\bar \pi ({\rm K}')} .} \label{eqtmp} \end{equation}

Particularly, when there is no degree dependence in the network, (\ref{eqtmp}) becomes
\begin{equation}{\bar \pi ({\rm K}) = {{\rm K}_{in}}\sum\limits_{{\rm K}'} {\frac{{{{{{{\rm K}'}_{out}}f_{{\rm K}'}} \mathord{\left/
 {\vphantom {{{{{\rm K}'}_{out}}f({\rm K}')} {\bar {\rm K}}_{in}}} \right.
 \kern-\nulldelimiterspace} {\bar {\rm K}}_{in}}}}{{{{{\rm K}'}_{out}}}}\bar \pi ({\rm K}')}  = \frac{1}{N}\frac{{{{\rm K}_{in}}}}{{\bar {\rm K}}_{in}} },\end{equation}

where ${\bar{\rm K}}_{in}$ is the average indegree in the network, implying that for networks with no degree-degree correlations, the RDS sample can be weighted by respondents' indegrees to estimate population proportions, which gives us the modified $VH_{out}$ estimator:

\begin{equation}
\hat p_A^{V{H_{in}}} = \frac{{\sum\limits_{{i} \in U \cap A} {{{(d_i^{in})}^{ - 1}}} }}{{\sum\limits_{{j} \in U} {{{(d_j^{in})}^{ - 1}}} }}.
\label{eq_RDS_in}\end{equation}

Clearly, the difference between $VH_{out}$ and $VH_{in}$ estimators is the substitution of outdegree with indegree. Thus, the use of $VH_{in}$ brings new challenges in the implementation of RDS as it requires collection of respondents' indegrees, which are not known from the RDS sample.

\subsection{Extension of $SH_{out}$ estimator to directed networks}
The $SH_{out}$ estimator was developed based on the fact that in any undirected network, the number of crossgroup edges pointing from $A$ to $B$, equals the number of edges pointing from $B$ to $A$. Similarly, in a directed network, the sum of nodes' indegrees in a group equals the total number of edges pointing to nodes in that group, i.e., if we let $S = \left[ {\begin{array}{*{20}{c}}
{{S_{AA}}}&{{S_{AB}}}\\
{{S_{BA}}}&{{S_{BB}}}
\end{array}} \right]$ be the recruitment matrix in the network, where, e.g., $S_{AB}$ is the proportion of edges originating in group $A$ which end in group $B$, then we have

\begin{equation}
\left\{ \begin{gathered}
  {N_A}\bar D_A^{out}{S_{AA}} + {N_B}\bar D_B^{out}{S_{BA}} = {N_A}\bar D_A^{in} \hfill \\
  {N_A}\bar D_A^{out}{S_{AB}} + {N_B}\bar D_B^{out}{S_{BB}} = {N_B}\bar D_B^{in} \hfill \\
\end{gathered},  \right.
\label{eq_degree_sum_formula}
\end{equation}

where, e.g., $\bar D_A^{out}$ is the average outdegree in group $A$.

For simplicity, let $m^* = \frac{{\bar D_A^{in}}}{{\bar D_B^{in}}}$ and $w^* = \frac{{\bar D_A^{out}}}{{\bar D_B^{out}}}$ be the average indegree and outdegree ratio of the two groups of nodes in the network, and let $\phi  = \frac{{{N_A}}}{{{N_B}}}$ be the relative group size proportion. Dividing the above equations (\ref{eq_degree_sum_formula}) gives a solution of $\phi$:

\begin{equation} \phi  = \frac{{w^*{{ S}_{AA}} - m^*{{ S}_{BB}}}}{{2m^*w^*{{ S}_{AB}}}} + \sqrt {\frac{{{{ S}_{BA}}}}{{m^*w^*{{ S}_{AB}}}} + {{(\frac{{m^*{{ S}_{BB}} - w^*{{ S}_{AA}}}}{{2m^*w^*{{ S}_{AB}}}})}^2}}.\label{eq_x4}\end{equation}
Then, if we can correctly estimate $m^*$, $w^*$ and $S$, we obtain a generalization of the $SH_{out}$ estimator:

\begin{equation}{\hat p_{A}^{SH_{in}}} = \frac{{\hat \phi }}{{1 + \hat \phi }},\label{eq_RDS_eq}
\end{equation}
in which we replace unknown population quantities in $\phi$ by their estimates from the RDS sample.

From the previous section, the average indegree ratio $m^*$ in $SH_{in}$ can be estimated by the harmonic mean ratio of indegrees from the sample for networks with no degree correlation: ${{\hat m}^*} = \frac{{{n_A}/\sum\limits_{{i} \in U \cap A} {{{(d_i^{in})}^{ - 1}}} }}{{{n_B}/\sum\limits_{{i} \in U \cap B} {{{(d_i^{in})}^{ - 1}}} }}$. It is however generally not possible to consistently estimate $w^*$ and $S$ using only the average outdegree and observed recruitment matrix. The sample mean outdegree will be an unbiased estimator only if there is no correlation between the indegree and outdegree of nodes, while the harmonic mean of outdegree is expected to have higher precision if the indegree-outdegree correlation is high. However, in simulations it is seen that there is little difference in using either the (arithmetic) mean or harmonic mean of outdegree to estimate $w^*$ and thereby we continue to use the harmonic mean in the following analysis. We have also tried to adjust potential bias in the estimation of $S$ by replacing individual inclusion probabilities with group inclusion probabilities (see Supportive Information (SI) for details), which however didn't improve the results and we therefore prefer to use the observed recruitment matrix from the sample to estimate $S$ in $SH_{in}$.

The factor $w^*$ was named the \textit{activity ratio} in \cite{Gile2010}, since it quantifies how active nodes in different groups are in building their personal networks. Following this, we henceforth refer to $m^*$ as the \textit{attractivity ratio}, as it reflects how ``attractive'' nodes in different groups are, or to which group of nodes edges are inclined to connect to.

\subsection{Sensitivity analysis when indegree is not known}

Hardly ever is the indegree observed in RDS studies. Consequently, the use of $VH_{in}$ and $SH_{in}$ is limited in practice. It is however possible to use both estimators if prior information is available. In $SH_{in}$, if the indegree is not known, the estimate of average indegree ratio, $\hat m ^*$, becomes an unknown parameter in (\ref{eq_RDS_eq}). This is true also for $VH_{in}$, since we can rewrite (\ref{eq_RDS_in}) as:
$${\hat p_{A}^{VH_{in}}} = \frac{{\sum\limits_{{i} \in U \cap A} {{{(d_i^{in})}^{ - 1}}} }}{{\sum\limits_{{j} \in U} {{{(d_j^{in})}^{ - 1}}} }} = \frac{{{n_A}/\hat {\bar D}_A^{in}}}{{{n_A}/\hat {\bar D}_A^{in} + {n_B}/\hat {\bar D}_B^{in}}} = \frac{{{n_A}/{n_B}}}{{{n_A}/{n_B} + \hat {\bar D}_A^{in}/\hat {\bar D}_B^{in}}}.$$
Replacing
${\hat {\bar D}_A^{in}}/{\hat {\bar D}_B^{in}}$ with $m$, we have:

\begin{equation}
{\hat p_{A}^{VH_m}} = \frac{n_A/n_B}{n_A/n_B+m}.\label{eq_indegM}
\end{equation}

Prior information may, for example, be obtained by expert opinion, or by using previous empirical results. What's more, even if there is little prior knowledge about the targeted population, we can, instead of providing a point estimate with fixed parameters, use a range of $m$ values to generate an estimate interval for $p_A^*$. That is, if $m^*$ is assumed to lie within a certain range, $[m_{min}, m_{max}]$, we get an interval of $\hat p_A$, $[\hat p_A(m_{min}), \hat p_A(m_{max})]$, by varying $m$ in (\ref{eq_RDS_eq}).

Following this, we will perform a sensitivity analysis on $SH_m$, and $VH_m$, by varying the ratio of average indegrees $m$ to get an interval of the estimates of $p_A^*$, $[\hat p_A(m_{min}), \hat p_A(m_{max})]$, with $m$ lying in a certain range, $[m_{min}, m_{max}]$. By choosing an interval centered on a value of $m$ based on prior information, we will get intervals of possible $\hat p_A$ values which more fully accounts for the situation when the network is directed, and provides valuable results on the sensitivity of estimators to the correctness of indegree assumptions about the network.

\section{Network Data and Study Design}

We will evaluate the performance of our suggested estimators and compare them with existing estimators through simulations of RDS processes on directed networks. The simulations will be performed on both artificially generated families of directed networks as well as a real MSM online social network~\cite{Xin2011}, which makes it possible to study the impact of different, carefully controlled, structural network properties on our estimators as well as looking at their behavior in a more realistic setting using actual data.

In our evaluation, we will consider the following parameters which are important both to directed networks and RDS estimation:

\textit{Directedness}; if $E_{dir}$ is the number of directed edges in a network with $E$ edges, then the proportion of directed edges is:
\begin{equation}\lambda  = {{{E_{dir}}} \mathord{\left/
 {\vphantom {{{E_{dir}}} E}} \right.
 \kern-\nulldelimiterspace} E},\label{eq_directedness}\end{equation}

i.e., $\lambda  = 0$ when the network is undirected, and $\lambda  = 1$ when the network is (extremely) directed in a way such that there are no reciprocal edges.

\textit{Indegree correlation}; the tendency that nodes with high indegrees are connected with each other. To quantify this, we use the assortativity ratio defined in~\cite{Newman2002}:

\begin{equation}\gamma  = \frac{{{E^{ - 1}}\sum\nolimits_i {{j_i}{k_i}}  - {{[{E^{ - 1}}\sum\nolimits_i {\frac{1}{2}({j_i} + {k_i})} ]}^2}}}{{{E^{ - 1}}\sum\nolimits_i {\frac{1}{2}(j_i^2 + k_i^2)}  - {{[{E^{ - 1}}\sum\nolimits_i {\frac{1}{2}({j_i} + {k_i})} ]}^2}}},
\label{eq_assortativity}
\end{equation}

where ${j_i}$ and ${k_i}$ are the indegrees of vertices at the end of the ${i^{th}}$ edge, $ i = 1, \ldots ,E$.

\textit{Indegree-outdegree correlation}; unlike the indegree correlation, which describes associations between nodes, the indegree-outdegree correlation measures the correlation between indegree and outdegree for the same node. We use the Pearson correlation calculated from all nodes in the network:

\begin{equation}\rho  = {{\operatorname{Cov} ({d^{in}},{d^{out}})} \mathord{\left/
 {\vphantom {{\operatorname{Cov} ({d^{in}},{d^{out}})} {{\sigma _{{d^{in}}}}{\sigma _{{d^{out}}}}}}} \right.
 \kern-\nulldelimiterspace} {{\sigma _{{d^{in}}}}{\sigma _{{d^{out}}}}}}.\label{eq_indegree_outdegree}
 \end{equation}

\textit{Homophily}; the probability that nodes connect with neighbors that are similar to themselves with respect to the studied feature $A$ rather than that they connect randomly~\cite{Morris1995,Rapoport1980,McPherson2001,Heckathorn2002}. Letting ${h_A}$ be the homophily for nodes with trait $A$, it holds that ${S_{AA}} = {h_A} + (1 - {h_A}){p_A}$, implying that ${h_A}$ can be calculated as:

\begin{equation}{h_A} = 1 - {{{S_{AB}}} \mathord{\left/
 {\vphantom {{{S_{AB}}} {{p_B}}}} \right.
 \kern-\nulldelimiterspace} {{p_B}}}.
 \label{eq_homophily}
 \end{equation}

The activity ratio $w^*$, as well as the attractivity ratio $m^*$, are also used as network structure parameters in our assessment.

\subsection{Network Data}

We will focus our study of network properties on directedness and attractivity ratio, as they are of general interest to the study of directed networks, and of particular interest to our estimators. Furthermore, we will vary other network structural properties that are likely to affect our estimators. For example, the $VH_{in}$ estimator is based on the assumption of no indegree correlation in the network, and the estimate of average outdegree in $SH_{in}$ is based on the assumption of positive indegree-outdegree correlation, etc.

In order to study the behavior of our estimators with respect to variation in directedness and attractivity ratio, we will use two families of generated networks, \textbf{Net1}, where there is little or no indegree correlation and no indegree-outdegree correlation, and \textbf{Net2}, which have varying homophily and positive indegree-outdegree correlation. This setting makes it possible to see how different structural properties, e.g. homophily, will affect our estimators as directedness and attractivity ratio are varied.

Net1 is generated starting from a random pure directed network, in which indegree and outdegree are uncorrelated ($\rho  \approx 0$). Then, the irreciprocal edges are rewired in a particular way that doesn't change nodes' degree in order to generate networks with different levels of directedness (down to $\lambda  = 0.2$) while the indegree-outdegree correlation remains unchanged. Finally, nodes are assigned either property $A$ or $B$ to achieve different attractivity ratios $m^* \in [0.7,{\text{ }}1.4]$ (see Table\,\ref{table 1}). The generating process for Net2 starts with a random undirected network. To obtain directedness, reciprocal edges are randomly rewired in such a way, that for any network in Net2 with directedness $\lambda $, the indegree-outdegree correlation is $\rho  \approx 1 - \lambda $. Then, different attractivity ratios are generated as for Net1, and we further rewire edges with respect to nodes' properties in order to achieve different levels of homophily: ${h_A} \in [0,0.5]$ (see Table\,\ref{table 1}). As we in this study are mostly interested in the case when sample size is relatively small compared to the population size, these networks are both of size 10,000.

\scalebox{0.8}{
\begin{threeparttable}[b]
\centering
\caption{\label{table 1}Basic statistics of Net1, Net2, Net3 and the MSM network}
\begin{tabular}
{@{\extracolsep{\fill}} cccccccccc}
\hline
\
& Network
& Average
& Directed-
& indegree corre-
& indegree-outdegree
& \
&Homophily
& Attractivity
&\ \\

\
& size ($N$)
& degree ($\bar D$)
& ness ($\lambda$)
& lation ($\gamma$)
& correlation ($\rho$)
& \
&($h$)
& ratio ($m^*$)
&$P$ \\
\hline

\textbf{Net1}&$10,000$&$10$&$[0,1]$&$[-0.09, 0.01]$&$\approx 0$&\ &$[-0.30,0.22]$&$[0.7,1.4]$&$70\%$\\
\hline
\textbf{Net2}&$10,000$&$10$&$[0,1]$&$[-0.03, 0.14]$&$\approx 1-\lambda$&\ &$[0,0.5]$&$[0.7,1.4]$&$30\%$\\
\hline

\ &\ &\ &\ &\ &\ &\textit{age}& $0.23$& $0.95$& $77\%$\\

\textbf{MSM} &$16,082$&$17.2$&$0.61$ &$0.03$ &$0.39$ &\textit{ct}& $0.50$& $1.32$& $39\%$\\

\textbf{Network} &\ &\ &\ &\ &\ &\textit{cs}& $0.03$& $0.96$& $40\%$\\

\ &\ &\ &\ &\ &\ &\textit{pf}& $0.06$& $1.05$& $38\%$\\
\hline
\textbf{Net3}&$--$\tnote{*}&$--$&$[0.61,0.91]$&$[0, 0.4]$&$--$&\ &$--$&$--$&$--$\\
\hline

\end{tabular}

\begin{tablenotes}
\item [*] {Same as the MSM network}
\end{tablenotes}

\end{threeparttable}

}

The anonymized online social \textbf{MSM network} used in this study (previously analyzed in \cite{Rybski2009, Xin2011}) has 16,082 nodes and comes from the Nordic region's largest and most active web community for homosexual, bisexual, transgender and queer persons (www.qruiser.com) and includes information on the relationships between members as well as members' personal information. Contacts between members on the web site are maintained by a ``favorites list'', on which each member can add any other member without approval from that member, so that the resulting social network will be directed. From this network, we obtain the giant strongly connected component of the friendship network of those members who identify themselves as homosexual males. Utilizing information from their user profiles, we can evaluate our estimators on different personal characteristics, and we will focus on four dichotomous properties: age (born before 1980), county (live in Stockholm, ct), civil status (married, cs), and profession (employed, pf). The proportions of nodes having a specific value of these properties are listed in Table\,\ref{table 1}.

While this network provides an opportunity to study our estimators in a more realistic setting, it and the studied personal properties will obviously have certain structural properties. In order to keep this level of realism, while still varying some structural network properties, based on the MSM network, we generate a family of networks, \textbf{Net3}, which have different levels of indegree correlation ($\gamma  \in [0,0.4]$). Detailed information on the generation process of  Net3 and the other networks can be found in the supportive information SI.

\subsection{Simulation Design}

RDS processes are then simulated on the above networks and estimates of RDS estimators are compared with true population properties. In each simulation, seeds are uniformly selected and coupons are randomly distributed to the recruiters' neighbors. To simulate RDS in real practice, we let the number of seeds be 10 and the number of distributed coupons be 3 when shorter sample waves are desirable, and, 6 and 2 for longer sample waves (provided in the supporting information). Sampling is done without replacement and we choose sample size 500 for Net1 and Net2, and 1000 for the MSM network and Net3. All simulations are repeated 1000 times.

For each simulation, we estimate the population proportion with our suggested estimators as well as existing estimators. Then, the root mean square error (RMSE), standard deviation (SD) and bias of estimators are calculated in order to quantify the results. The estimators are divided into four categories:

(i) The na\"{i}ve estimator: The raw sample composition;

(ii) Outdegree-based estimators: $SH_{out}$ and $VH_{out}$;

(iii) Indegree-based estimators: $SH_{in}$ and $VH_{in}$;

(iv) Estimators based on known parameters: $SH_{m^*}$ and $VH_{m^*}$.

Note that the indegree-based estimators are practically useless, since individual indegrees are not known from the RDS sample, and are therefore presented merely for comparison and theoretical purposes.

Additionally, we include the estimator ${SS}$ (Succesive Sampling) suggested in~\cite{Gile2011} for reference; note that this estimator is based on a different estimation procedure and requires knowledge of the population size ($N$) in order to yield correct estimates. Since the $SS$ estimator is developed for RDS on undirected networks, two versions of it will be used in order to adapt it for use with directed networks: $SS_{out}$ using outdegrees of respondents in the sample in the estimation procedure and $SS_{in}$ using their indegrees. In the estimation procedure of $SS_{out}$ and $SS_{in}$, $M=500$ times successive sampling samples per each of $r=3$ iterations are used.

\section{Results}

\subsection{Estimation performance on networks with known properties}

\subsubsection{Networks with varying structural properties}

\begin{figure*}
 \centerline{\includegraphics[width=.8\textwidth]{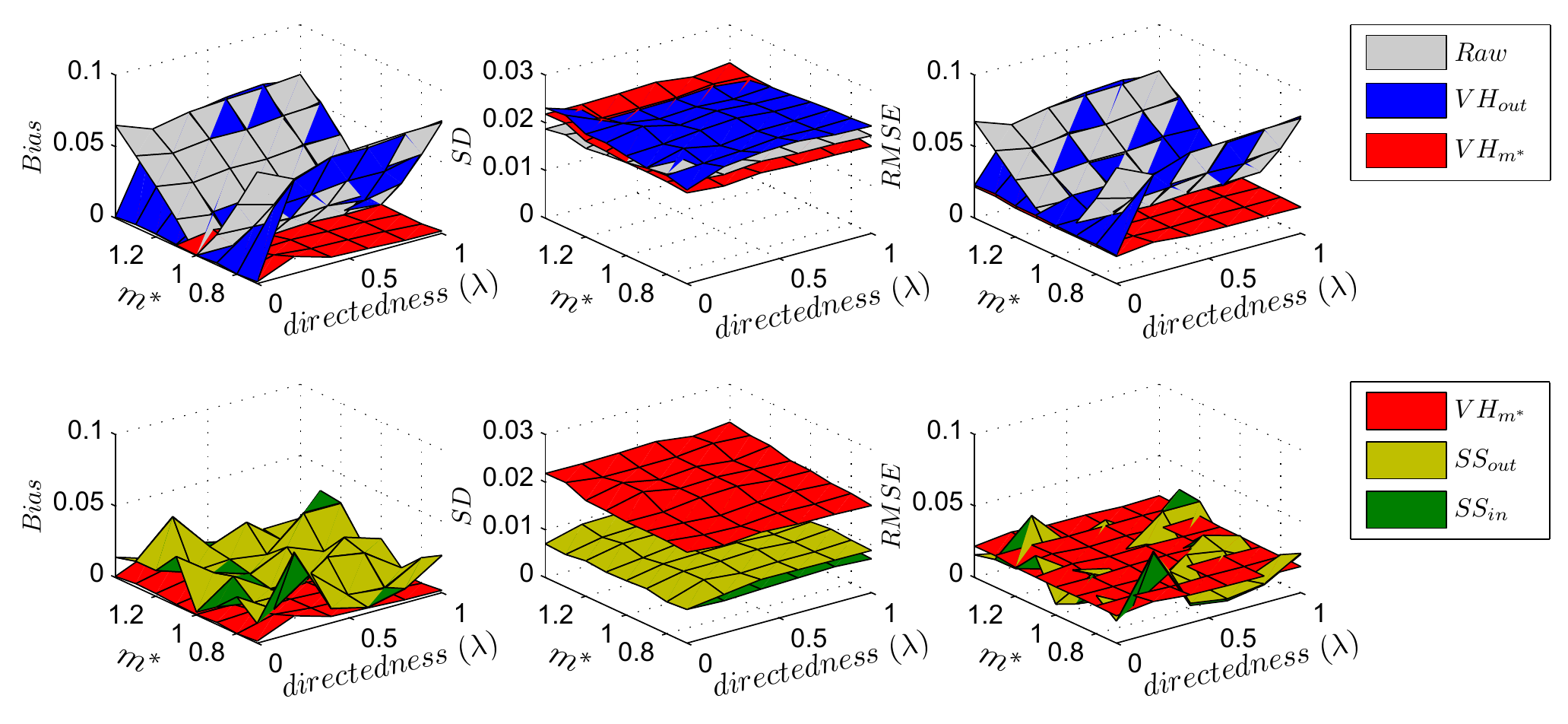}}
 \caption{Bias, Standard Deviation, and Root Mean Square Error of RDS estimators on Net1. Sampling without replacement, number of seeds=10, coupons=3, sample size=500.}\label{fig1}
 \end{figure*}

We start by looking at how varying directedness and attractivity ratio affects RDS estimators in an otherwise uncomplicated setting, i.e., the generated networks with close to zero indegree correlation and no indegree-outdegree correlation, \textbf{Net1}. In Figure \ref{fig1}, the bias, SD and RMSE of the raw sample composition, $VH_{out}$ and $VH_{m^*}$ are shown in the top row ($SH_{in}$, $SH_{m^*}$, and $VH_{in}$ perform very similar to $VH_{m^*}$ and are thus left out), and the same figures for $VH_{m^*}$, $SS_{out}$, and $SS_{in}$ are shown in the bottom row for visual clarity.

We can see in the top row that both the raw sample composition and $VH_{out}$ are biased with increasing $|m^*-1|$, and that $VH_{out}$ has the same level of bias and RMSE as the sample composition as long as the network is directed, i.e., $\lambda>0$. On the other side, the indegree-based estimator, $VH_{m^*}$, generates negligible bias and consistently smaller RMSE.

In the bottom figures, we see that $SS_{in}$ and $SS_{out}$ have varying bias (smaller than $VH_{out}$) as directedness and attractivity ratio changes, while retaining a substantially lower SD; on the other hand, $VH_{m^*}$ has smaller bias but larger SD. Consequently, the RMSE of $SS_{in}$ and $SS_{out}$ becomes similar with that of $VH_{m^*}$ due to their small SD; sometimes, the RMSE of $SS_{in}$ and $SS_{out}$ is even smaller than that of $VH_{m^*}$.

\begin{figure*}
 \centerline{\includegraphics[width=.8\textwidth]{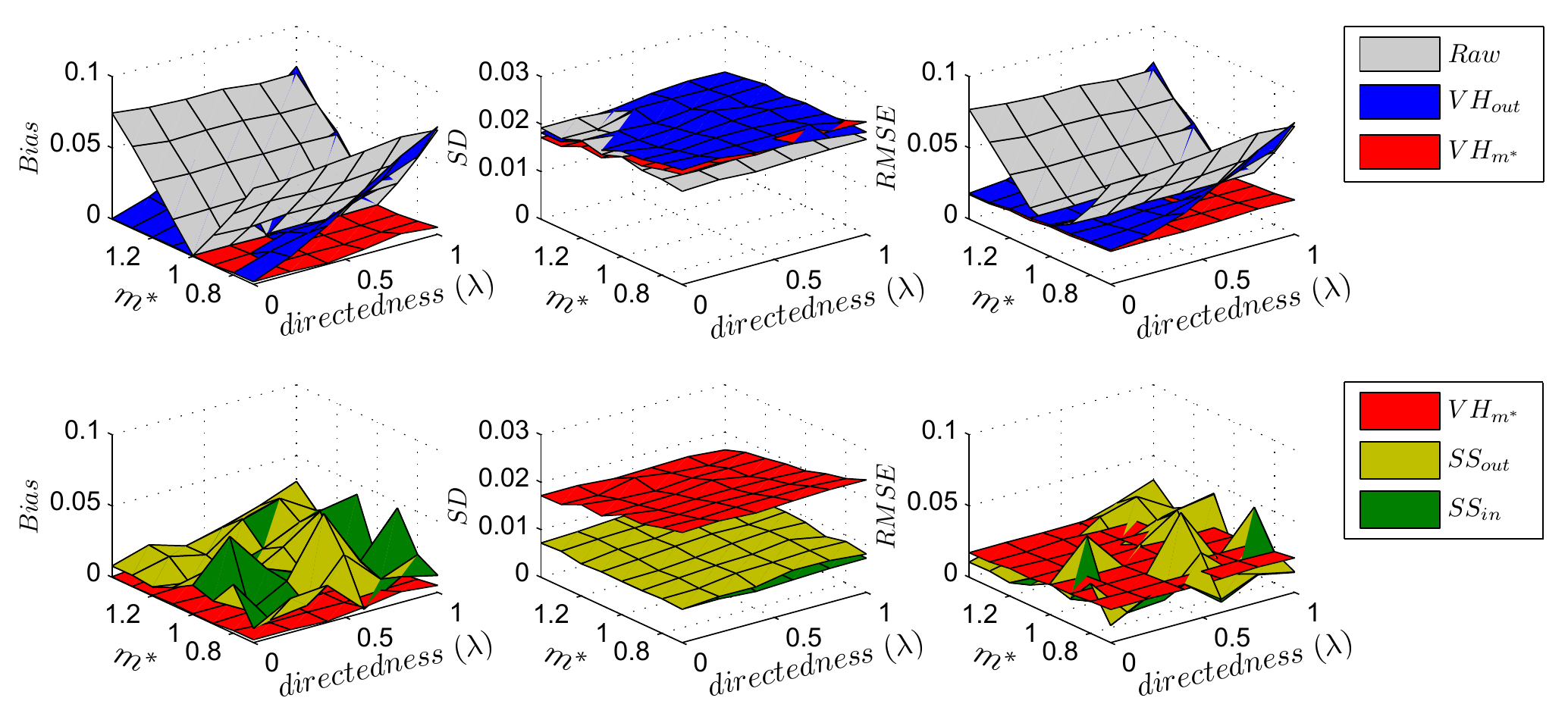}}
 \caption{Bias, Standard Deviation, and Root Mean Square Error of RDS estimators on Net2, homophily ${h_A} = 0$. Sampling without replacement, number of seeds=10, coupons=3, sample size=500.}\label{fig2}
 \end{figure*}

The results in Figure \ref{fig2} for RDS on networks with varying indegree-outdegree correlation, but no homophily, \textbf{Net2}, are similar to those seen in Figure \ref{fig1}, except that the bias and RMSE of $VH_{out}$ now increase gradually with the directedness of the network, generating bias and RMSE smaller than the raw sample composition, but larger than $VH_{m^*}$. This difference in performance of $VH_{out}$ on Net1 and Net2 shows that for RDS on network with indegree-outdegree correlation, the traditional outdegree-based estimators, can still be expected to give less estimate bias and error than the raw sample composition. However, the indegree-outdegree correlation have little effect on the performance of estimators utilizing known parameters, i.e., $SS_{in}$, $SS_{out}$, and $VH_{m^*}$.

\begin{figure*}
 \centerline{\includegraphics[width=.8\textwidth]{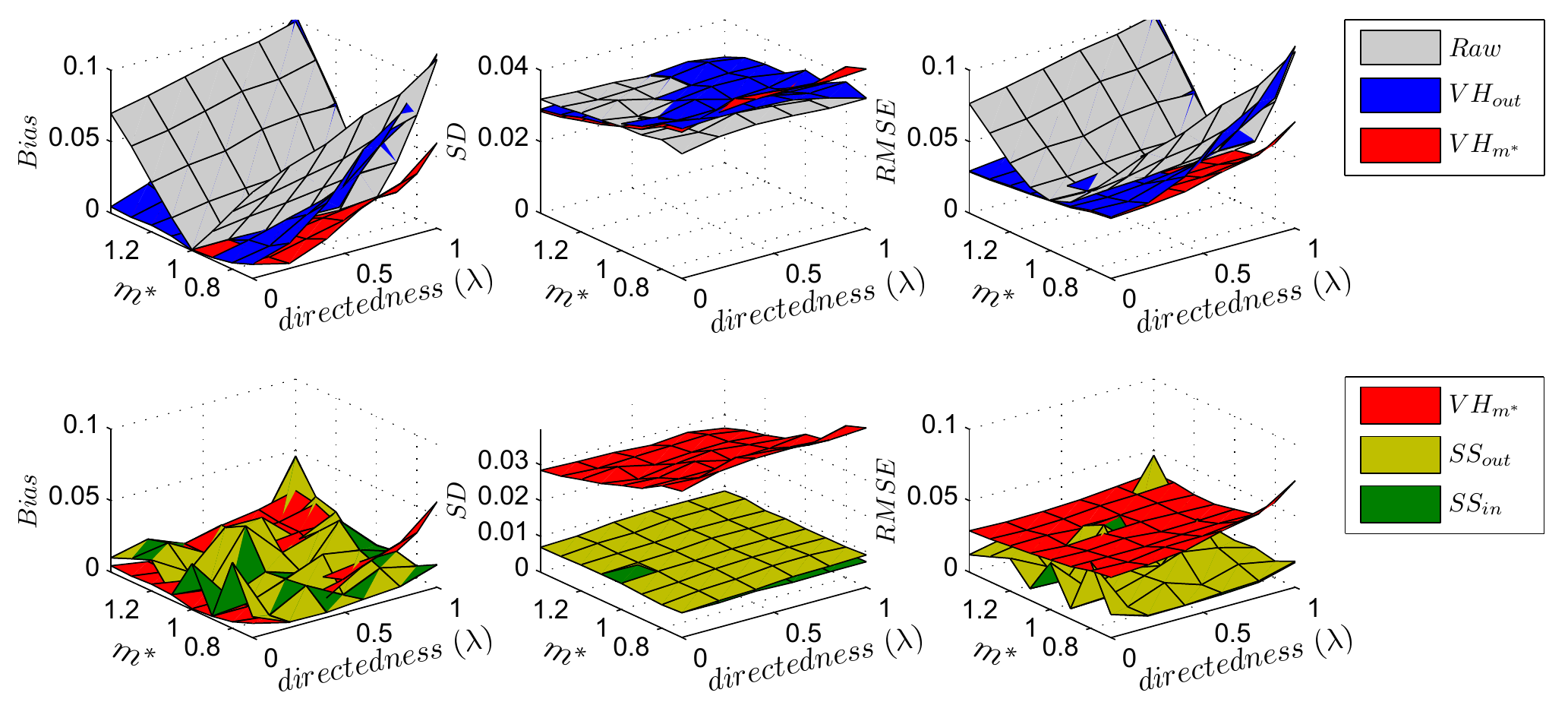}}
 \caption{Bias, Standard Deviation, and Root Mean Square Error of RDS estimators on Net2, homophily ${h_A} = 0.4$. Sampling without replacement, number of seeds=10, coupons=3, sample size=500.}\label{fig3}
 \end{figure*}

In Figure \ref{fig3}, where homophily $h=0.4$, we see that the magnitude of bias, SD and RMSE all increase for the raw sample composition, $VH_{out}$ and $VH_{m^*}$, indicating a clear effect of homophily on increasing RDS estimate bias and error. However, on the other hand, $SS_{in}$ and $SS_{out}$ are quit robust to the effect of homophily, their SD ($[0.008, 0.01]$) remains substantially smaller than the rest estimators ($[0.02, 0.04]$), and in most cases they produce minimum RMSE.

Overall, it is clear that previous RDS estimators are seriously affected by letting RDS processes take place on directed networks, and that our suggested estimators and the $SS$ estimators, although all relying on previous knowledge about the network, shows major improvements in the quality of estimates.

\subsubsection{The MSM network and its modifications}

\begin{figure*}
 \begin{center}
 \centerline{\includegraphics[width=0.82\textwidth]{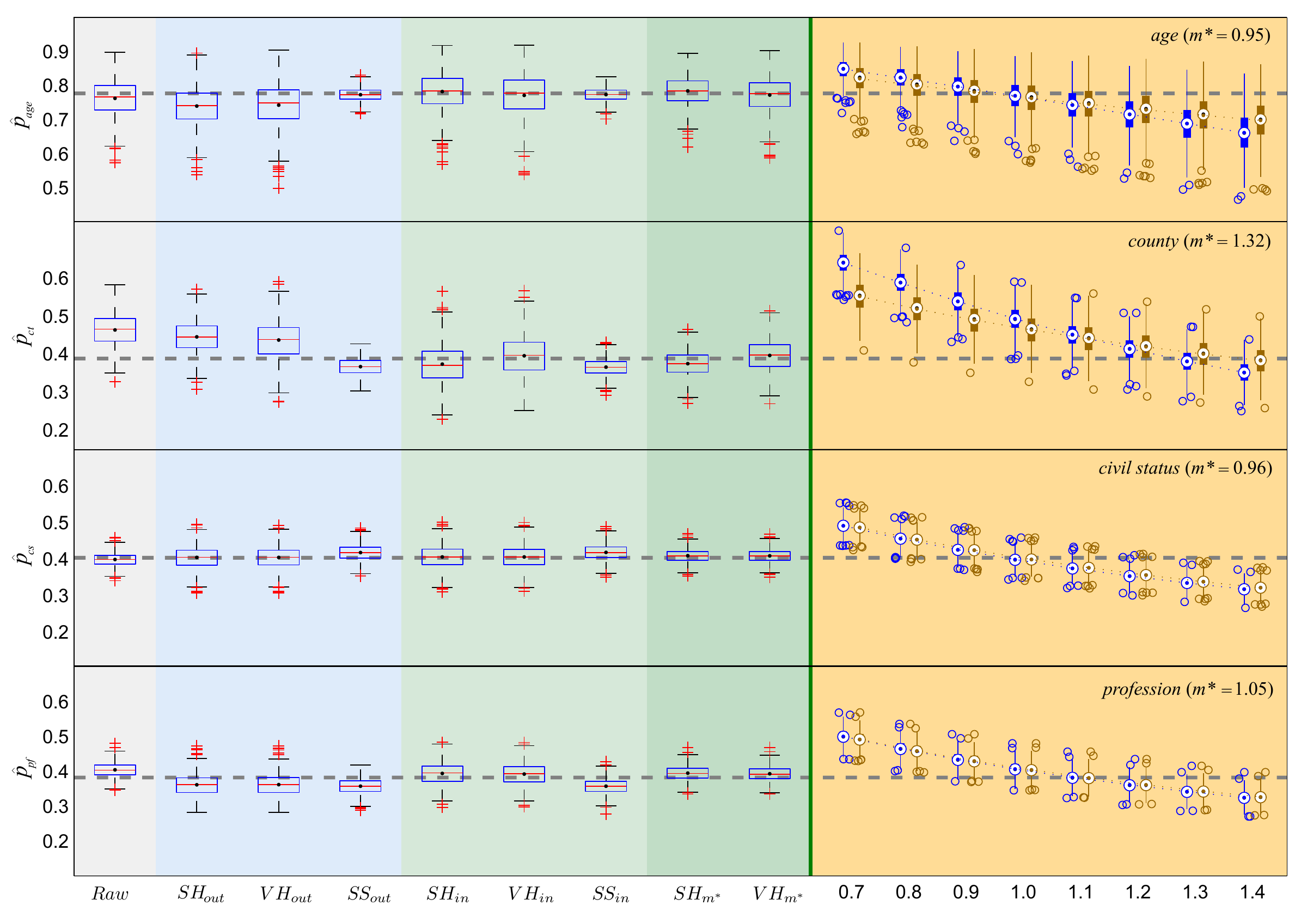}}
 \caption{RDS on MSM network. The right panel shows sensitivity analysis of $\hat p_{A}^{VH_m}$ (brown) and $\hat p_{A}^{SH_m}$  (blue) with $m$ varying from 0.7 to 1.4, plots are horizontally shifted a few points to avoid overlapping. Sampling with replacement, number of seeds=10, number of coupons=3, sample size=1000.}\label{fig4}
 \end{center}
 \end{figure*}

In the \textbf{MSM network} we look at four dichotomous user properties and how the estimators behave on each of them. As the structural properties of the MSM network are fixed, we illustrate estimator behavior through box plots, which are shown in the left panel of Figure \ref{fig4}. In each box, the central line is the median, the dot is the mean, the edges of the box are the 25th ($q_1$) and 75th ($q_3$) percentiles. Estimates being at least $1.5(q_3-q_1)$ away from the edges of the box are shown as outliers beyond the whiskers.

The traditional outdegree-based estimators, $SH_{out}$ and $VH_{out}$, have large bias when estimating variables with large homophily and attractivity ratios which significantly differ from 1, i.e., age and county. For example, their estimates of the proportion of MSM members who live in Stockholm are on average over 5 percentage units higher than the true value, and for age, civil status and profession, the sample mean has even less bias than them.

The indegree-based estimators, $SH_{in}$ and $VH_{in}$, are generally much less biased for all variables, indicating that the indegree is a good approximation of sampling probability for nodes in directed networks. The $m^*$-based estimators, $SH_{m^*}$ and $VH_{m^*}$, have a similar performance on the biasedness, with slightly smaller SD.

Lastly, when we look at the $N$-based estimators, $SS_{out}$ and $SS_{in}$, they are biased for all the four variables, however, the (substantially) smaller SD makes their error lie within an acceptable range compared to $SH_{out}$ and $VH_{out}$.

\begin{figure*}
 \begin{center}
 \centerline{\includegraphics[width=0.82\textwidth]{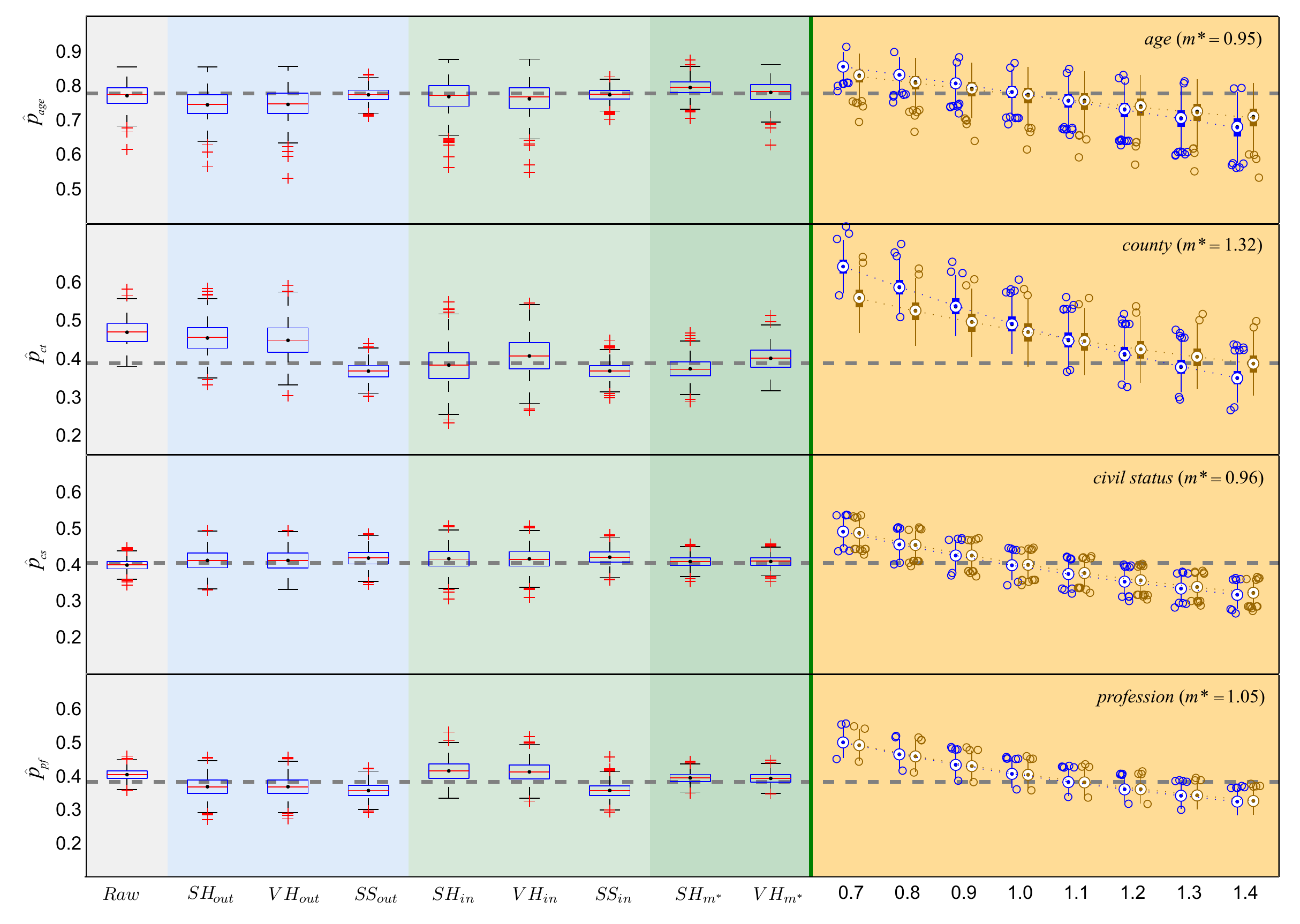}}
 \caption{RDS on Net3 with indegree correlation $\gamma  = 0.4$. The right panel shows sensitivity analysis of $\hat p_{A}^{VH_m}$ (brown) and $\hat p_{A}^{SH_m}$ (blue) with $m$ varying from 0.7 to 1.4, plots are horizontally shifted a few points to avoid overlapping. Sampling without replacement, number of seeds=10, number of coupons=3, sample size=1000.}\label{fig5}
 \end{center}
 \end{figure*}

In Figure \ref{fig5}, we can see that the results from simulations on the modified MSM network, \textbf{Net3} with indegree correlation $\gamma=0.4$, are very similar to the results from the unmodified MSM network. The indegree correlation gives the indegree-based estimators, $SH_{in}$ and $VH_{in}$, together with the $m^*$-based estimators, $SH_{m^*}$ and $VH_{m^*}$, a slightly increased bias, however, the overall performance of these estimators are better than $SH_{out}$ and $VH_{out}$.

The results from the two $SS$ estimators are practically unchanged from the MSM network, which indicates that these estimators are very robust to changes in indegree correlation. Again we find  (substantially) smaller SD over all variables generated by these estimators, which makes the estimated error of $SS_{out}$ and $SS_{in}$ comparatively small despite that they are biased on directed networks.

\subsection{Sensitivity analysis}
\begin{figure*}
 \begin{center}
 \centerline{\includegraphics[width=.8\textwidth]{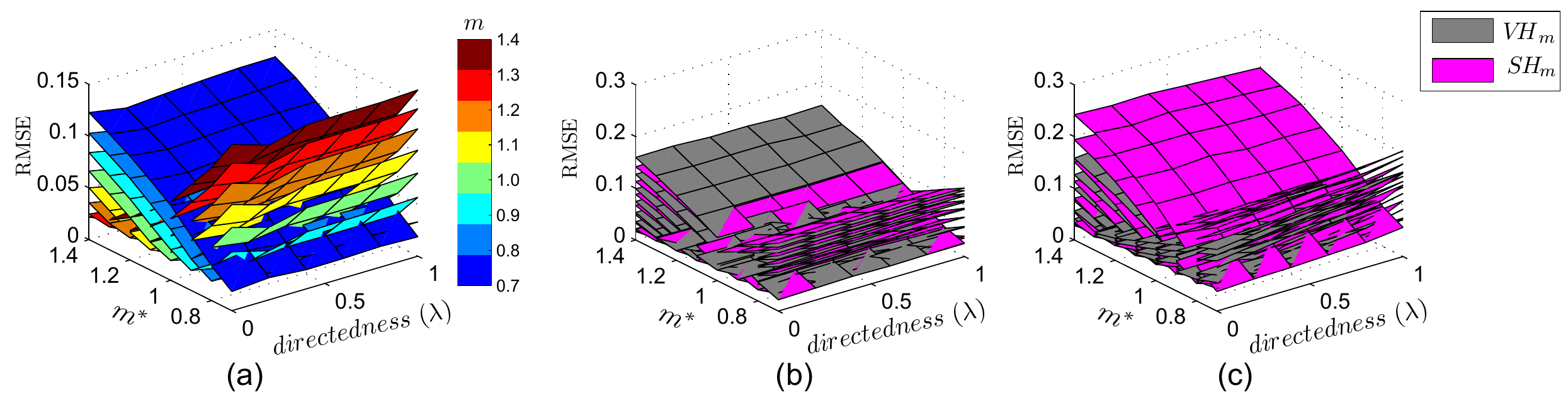}}
 \caption{Sensitivity analysis of $\hat p_{A}^{VH_m}$ and $\hat p_{A}^{SH_m}$ on Net1 and Net2 with tested $m$ values. (a) Net1, $\hat p_{A}^{SH_m}$ not shown as it is similar to $\hat p_{A}^{VH_m}$; (b) Net2 with homophily ${h_A} = 0$; (c) Net2 with homophily ${h_A} = 0.4$. Sampling without replacement, number of seeds=10, coupons=3, sample size=500.}\label{fig6}
 \end{center}
 \end{figure*}

We perform sensitvity analysis on $SH_m$ and $VH_m$ with respect to the attractivity ratio $m$. The results from the generated networks can be seen in Figure \ref{fig6}.

Figure \ref{fig6}(a) shows how the RMSE of $VH_m$ changes with directedness and attractivity ratio when different values of $m$ are given to the estimator as simulations are performed on \textbf{Net1}. It is clear that proper prior information on $m$ will give small error in the estimator; note also that changes in directedness does not affect estimator performance.

Figures \ref{fig6}(b) and \ref{fig6}(c) shows the RMSE of $VH_m$ and $SH_m$ from simulations on \textbf{Net2} with homophily $h=0$ and $h=0.4$ respectively, and it can be seen that while the estimators have similar performance when homophily is low, $VH_m$ generate less RMSE when $m$ is far away from $m^*$ and homophily is high, implying that when $m^*$ is not known, $VH_{m}$ may be a better option than $SH_{m}$ in real practice.

In the sensitivity analysis on the \textbf{MSM network} and \textbf{Net3}, which can be seen on the right side of Figures \ref{fig4} and \ref{fig5}, there is more variability in the estimates from age and county than from profession and civil status, as would be expected from the previous results. We see that the change in $VH_m$ is smaller than in $SH_m$ as $m$ is varied; this is however negligible for profession and civil status. Generally, we see that we will cover the true value well by using the average estimates from the sensitivity analysis, which is especially interesting for county, the only property of which $m^*$ significantly differs from 1.

Overall, we can see that $VH_m$ performs better than $SH_m$, making it the preferred choice for RDS in real practice. We also did simulations on the above networks with 6 seeds and 2 coupons; however, no substantial differences are found in the results on the performance of estimators nor for the sensitivity analysis, see supplementary material.

\subsection{Confidence interval and implementation}
An problem associated with the use of $VH_m$, is to construct a confidence interval around $\hat p_A^{VH_m}$ when $m^*=m$. Traditionally, the standard error of RDS estimates are generated by a bootstrap procedure~\cite{Salganik2006}, in which replicated samples are drawn based on the recruitment property of original RDS samples. We modify the traditional bootstrap method by letting $\hat p_A^{VH_m}$ substitute the traditional RDS estimator when each bootstrapped sample is produced, and then let the middle 90\% (95\%) of the ordered estimates from the bootstrap samples' estimates be the approximation of the confidence interval.

\begin{figure*}[ht]
 \begin{center}
 \centerline{\includegraphics[width=1\textwidth]{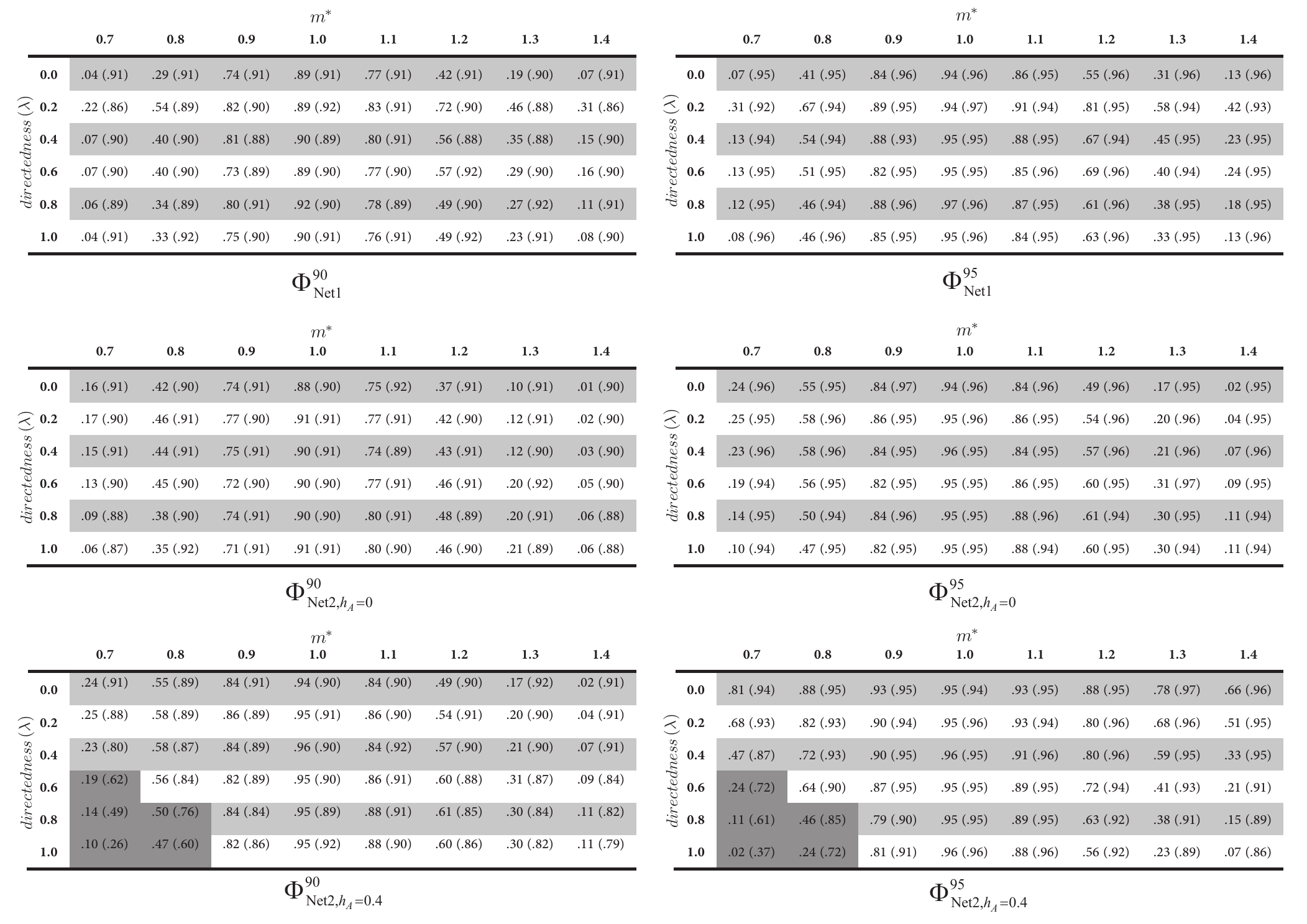}}
 \caption{90\% and 95\% bootstrap coverage probability of $\hat p_{A}^{VH_{out}}$ and $\hat p_{A}^{VH_{m^*}}$ (shown in brackets) on Net1 and Net2. Sampling without replacement, number of seeds=10, coupons=3, sample size=500.}\label{fig7}
 \end{center}
 \end{figure*}

We test the above procedure on Net1 and Net2; for each simulation setting ($[\lambda, m^*, h_A]$), we take 1000 RDS samples and for each of these 1000 samples we construct 90\% and 95\% confidence intervals based on 1000 replicate samples drawn by the above bootstrap procedure. The proportion of times that the generated confidence interval contains the true population value $p_A^*$, denoted as $\Phi^{90}$ and $\Phi^{95}$, are compared with the coverage rates of the traditional RDS estimator based method and are presented in Figure~\ref{fig7}.

Apparently, due to the large bias of $VH_{out}$ when network directedness and attractivity ratio is high, the traditional bootstrap procedure performs quite poorly with respect to $\Phi^{90}$ and $\Phi^{95}$. The attractivity ratio has substantial impact; when $m^*=1$, the coverage rate is generally close to the desired confidence level; however, the coverage rate drops rapidly as $m^*$ deviates away from 1. For example, when $m^*=0.8$, the coverage rate of the 90\% confidence interval is only 29\% even when the network (Net1) is undirected.

On the other hand, the $VH_m$-based bootstrap procedure gives reliable and consistent confidence intervals over all network settings. With the exception of some extreme cases, i.e., when $m^* \le 0.9$, $\lambda >0.6$, and $h_A=0.4$, the coverage rate is fairly close to the desired confidence level and are overall better than that of the $VH_{out}$-based procedure.

It is of interest to use the previously suggested sensitivity analysis together with the given bootstrap procedure. As an illustration on how to implement the proposed methods in real RDS practice, when indegree information is not collected, we take data given in~\cite{Abdul-Quader2006} and perform sensitivity analysis with $VH_{m}$, providing confidence intervals for all values of $m$. A sample of 618 drug users in New York City, and their personal characteristics, were collected using RDS with eight seeds. By using our methods on this data, we produce estimates and 90\% confidence intervals on the proportion of males and the proportion of injectors.

It is not obvious which values of $m$ that should be used in the sensitvity analysis. One suggestion is to let $m$ vary around the observed activity ratio $\hat w^*$, since the indegree-outdegree correlation is positive in most social networks~\cite{South2004, Wallace1966, Feld2002}. The activity ratio ($\hat w^*$, weighted) for males is 0.99, indicating that there is little difference of the size of personal networks with respect to gender. However, the activity ratio for injectors is 1.58, indicating that injecting drug users know substantially many more drug users than those who don't inject drugs. The length of the interval of $m$ is arbitrarily set to 1.

\begin{figure*}[ht]
 \begin{center}
 \centerline{\includegraphics[width=.8\textwidth]{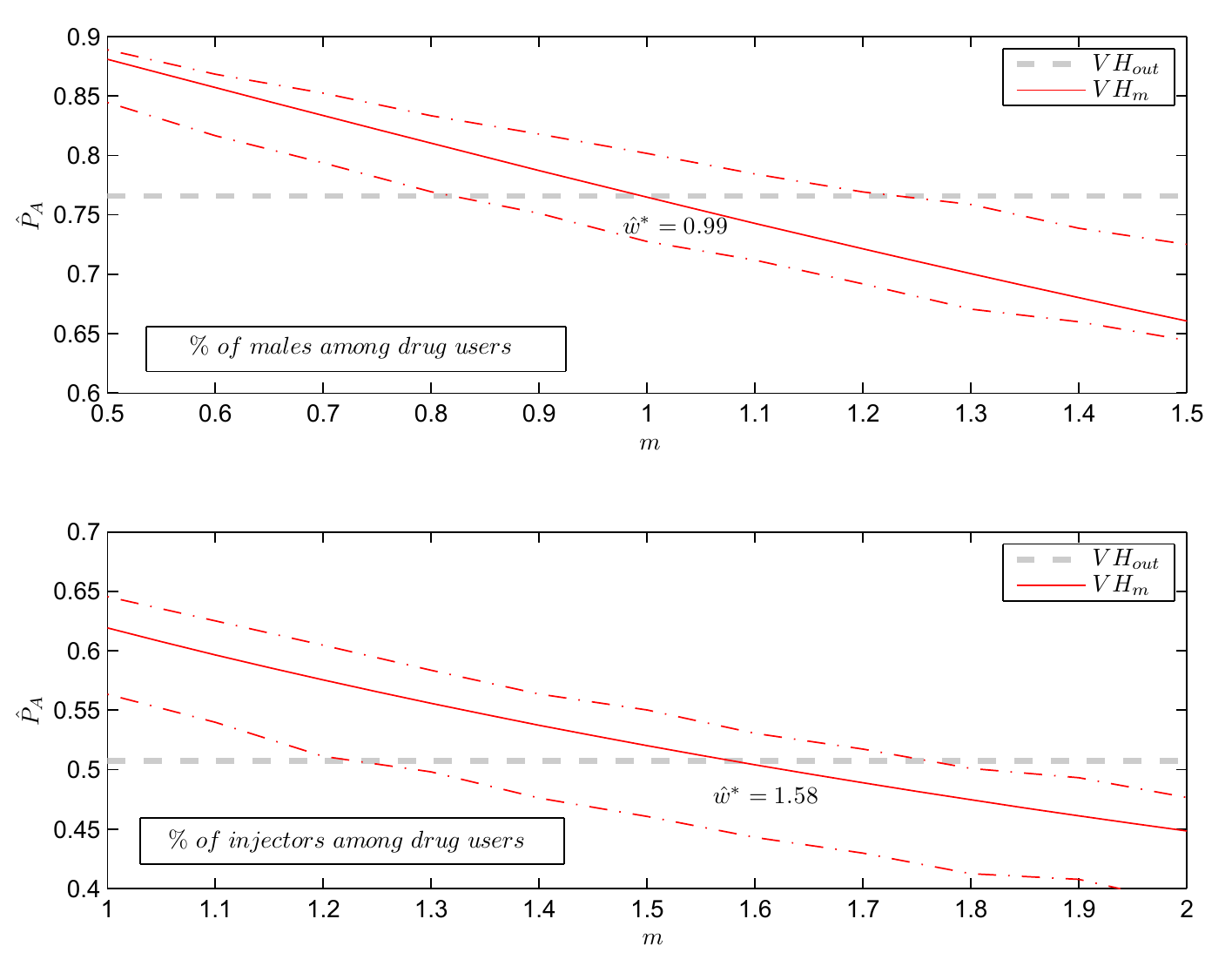}}
 \caption{Sensitivity analysis of RDS estimates for proportion of (a) males and (b) injectors among drug users in New York City. The dash-dot line shows 90\% CI of $\hat p_{A}^{VH_m}$.}\label{fig8}
 \end{center}
 \end{figure*}

In Fig.\,\ref{fig8}, we can see that when $m = \hat w^*$, the $VH_{m}$ estimates are equal to those given by $VH_{out}$. When the network is assumed directed and $m \in [0.5,{\rm{ }}1.5]$, the estimated proportion of male drug users will vary from 0.88 to 0.66. The proportion of injecting drug users, varies from 0.45 to 0.62 when $m \in [1,{\rm{ }}2]$. The $m$ intervals used here are arbitrarily chosen and their precision thus unknown, and therefore, it is hard to draw major conclusions from this example. However, the above analysis conveys another important information: for each change of 0.1 in the average indegree ratio, the change in the RDS estimates will be about 2 percentage units, which also may be an indication of how sensitive the RDS estimates are to uncertainties in the collected degree data.

\section{Conclusion and Discussion}

Despite the widely acknowledged evidence of the existence of directedness among social networks, the effect of directedness on RDS estimates has seldom been evaluated. This could be problematic since all previously reported RDS estimates rely on the assumption that the studied networks are purely reciprocal, the violation of which will result in unknown bias. To address this situation, we have extended previous RDS estimators onto directed networks and evaluated their performance on networks with various structural properties.

Our study shows that the individual indegree is a fair approximation to nodes' sampling probabilities in a RDS process on a directed network, and that this approximation is robust to changes in indegree-outdegree correlation and indegree correlation etc. The suggested indegree-based estimators $SH_m$ and $VH_m$ can also be used with prior information on $m^*$, the ratio of harmonic mean indegrees of the two groups in the sample; an approach that gives good results. Prior information about $m^*$ may possibly be obtained by expert opinions, or by using previous empirical studies related to the studied population; the need for prior information however limits this application in practice.

To further address the reality that the indegrees are not observed from an RDS sample, we have developed a sensitivity analysis method, based on the attractivity ratio $m^*$ between the studied groups in the population, to incorporate the uncertainties in both network directedness and reported outdegrees. Our results show that, while it is of course best to have correct prior information on the network, it is possible to get a deeper understanding of how RDS estimation is influenced by network directedness by using sensitivity analysis. E.g., in our results from the MSM network it is shown that erroneous prior information may give serious errors.

It is difficult to find an obvious setup for the sensitivity analysis both with respect to guessing a value of $m^*$ and deciding to which extent it should be varied. However, since many social networks have positive indegree-outdegree correlation, the activity ratio $\hat w^*$, which is observed from the sample, may be an indicator of where to vary $m$ from, and often the difference between $m^*$ and $w^*$ is small. For example, in the MSM network, the absolute difference is 0.27, 0.17, 0.02, and 0.05 for age, county, civil status and profession, respectively~\cite{Xin2011}.

From the results of sensitivity analysis on Net1 and Net2, we can see that the performance of $VH_{m}$ and $SH_{m}$ is determined primarily by the attractivity ratio $m^*$, rather than network directedness $\lambda$. Thus, if the network instead is assumed undirected, in which the ratio of indegrees is equal to the ratio of outdegrees ($m^*=w^*$), the sensitivity analysis may instead be used to assess the uncertainty of reported (out)degrees. The differential function of $VH_m$ over $m$, $\frac{{\partial V{H_m}}}{{\partial m}}\left| {_{m = {{\hat w}^*}}} \right.{\rm{ = }}{(\frac{{\frac{{{n_A}}}{{{n_B}}}}}{{\frac{{{n_A}}}{{{n_B}}} + m}})^\prime }\left| {_{m = {{\hat w}^*}}} \right. =  - \frac{{\frac{{{n_A}}}{{{n_B}}}}}{{{{(\frac{{{n_A}}}{{{n_B}}} + {{\hat w}^*})}^2}}}$, then provides the magnitude of how much the RDS estimate would change if there is any reporting error in the degree information.

Acting as a natural companion to the sensitivity analysis or estimation with prior information, our modification of the previous RDS bootstrap method has proven to work well on directed networks. Based on the $VH_m$ estimator, it largely outperforms the traditional $VH_{out}$-based method, and can construct close-to-expected confidence intervals for networks with varying directedness and attractivity ratio.

Another finding, which has not been highlighted in previous research, concerns the $SS$ estimator~\cite{Gile2011}. This estimator has small and overall consistent standard error among the networks tested in our paper. Given that the population size is known, this estimator is expected to produce RDS estimates with acceptable bias and error.

While it is in the interest of this study to do a full evaluation of RDS on directed networks, it is worth noting that since RDS utilizes a peer-driven mechanism, and the recruitment rights are few and valuable, respondents are usually inclined to recruit those that they know reasonably well. Such a mechanism to a large extent avoids the occurrence of recruitment via directed edges, and in most RDS studies, the proportion of recruitment through strangers are relatively small, usually less than 10\%~\cite{Wejnert2008, Iguchi2009, Ma2007, Dana2011, Daniela2009}; as pointed out earlier, this proportion may in some cases be larger though. In our results, we see that already small proportions of directed edges affects previous estimators, so while networks with extremely high directedness is very unlikely to occur in reality, and primarily are included out of theoretical interests, our evaluation shows that estimators will be sensitive to directedness in reality and that it is therefore an important issue to address.

For actual RDS practice, network directedness has previously not been an highlighted issue. The suggested sensitivity analysis gives RDS practitioners the possibility to take directedness into account as it provides means to understand the robustness of sample inference to the violation of certain assumptions: that the network may be partially directed, and that the degree data collected from respondents may contain reporting error. As there are no methods available on how to quantify network directedness, an interval of estimates based on a range of $m$ values is currently the best way of understanding this issue; additionally, it may give researchers a more detailed image of the situation and advice on how to understand the studied population.

We hope that the current study can inspire research beyond the purpose of studying hidden populations, such as sampling the contents of webpages, where indegree may be used as a cheap and efficient parameter to approximate the inclusion probability of random walks on internet~\cite{Fortunato2007, Neill2001, Chareen2005, Gjoka2010}. Other factors that might affect inclusion probabilities, such as transitivity, degree distribution, closeness, etc., yet need to be investigated in future studies.

\section*{Acknowledgments}
This work is funded in part by Riksbankens Jubileumsfond (dnr: P2008-0674) and the Swedish Research Council. X.L. would like to thank China Scholarship Council (Grant No. 2008611091). Thanks are also due to Sida for their support to RDS development work in Vietnam.


\section*{Appendix A: Generation process of Net1, Net2 and Net3}
\subsection*{A.1. Net1}\label{sectionNet1}
Net1 is the set of networks with different levels of directedness, in which the indegree and outdegree are not correlated ($\rho \approx 0$).

\textbf{Step 1} \textit{Base network}. At first, a random purely directed network ($\lambda=1$) is generated by randomly distributing $N\bar{D}$ irreciprocal/directed edges between $N$ nodes with the restriction that no reciprocal edges should be formed.

\begin{figure*}[ht]
\centerline{\includegraphics[width=.3\textwidth]{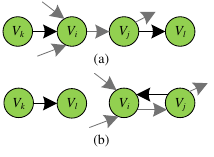}}
\caption{Net1: Illustration of the rewiring process leading to a decrease in $\lambda$}\label{sup_fig1}
\end{figure*}

\textbf{Step 2} \textit{Varying directedness}. In order to decrease the directedness to a given $\lambda  \in [0.2, 1]$, irreciprocal edges in the base network are randomly chosen and rewired to form reciprocal edges. Specifically, at each step, an edge $i \to j$ between nodes $v_i$ and $v_j$ is randomly chosen; then, if there is no link pointing from $j$ to $i$, we randomly find an irreciprocal incoming edge of $i$, ${k \to i}$ and an irreciprocal outgoing edge of $j$, ${j \to l}$ (\autoref{sup_fig1}(a)). These edges are then rewired as $k \to l$, and $j \to i$ (\autoref{sup_fig1}(b)), such that a new reciprocal pair of edges $i \leftrightarrow j$ is formed, and the degrees of $i$, $j$, $k$ and $l$ remain unchanged. The rewiring process is restarted from the beginning if the network is disconnected.

\textbf{Step 3} \textit{Varying attractivity ratio}. Let ${m^*}$ be the desired attractivity ratio. For each network generated in \textbf{Step 2}, $NP^*$ ($0<P^*<1$) nodes are randomly picked and assigned property $A$, and the remaining nodes are assigned property $B$. Then, the attractivity ratio of the network, $m'=\frac{\bar{d}_A^{in}}{\bar{d}_B^{in}}$, where $\bar{d}_A^{in}$, $\bar{d}_B^{in}$ are the average indegrees of nodes with property $A$ and $B$ respectively, is calculated. If $m' \ne {m^*}$, the following algorithm is carried out in order to generate a network with the required $m^*$ value:

(i) Randomly pick two nodes, $v_i$ and $v_j$, with different properties;

(ii) If $m' > m^*$ and $\bar{d}_A^{in} > \bar{d}_B^{in}$, exchange the property of $v_i$ and $v_j$;

(iii) Else if $m' < m^*$ and $d_A^{in} < d_B^{in}$, exchange the property of $v_i$ and $v_j$;

(iv) Repeat (i)-(iii) until $m'=m^*$.

\textbf{Step 4} A random undirected network of the same size and average degree is generated separately for $\lambda = 0$, and the method described in \textbf{Step 3} is used to generate different $m^*$ values in this network.

\subsection*{A.2. Net2}
Net2 is the set of networks with a certain amount of indegree-outdegree correlation and different levels of directedness and homophily.

\textbf{Step 1} \textit{Base network}. At first, a random purely undirected network ($\lambda=0$) is generated by randomly distributing $N\bar{D}/2$ reciprocal/undirected pairs of edges between $N$ nodes.

\begin{figure*}[ht]
\centerline{\includegraphics[width=.3\textwidth]{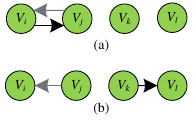}}
\caption{Net2: Illustration of the rewire process carried out to increase $\lambda$}\label{sup_fig2}
\end{figure*}

\textbf{Step 2} \textit{Varying directedness}. In order to increase the directedness to a given $\lambda \in [0, 1]$, reciprocal edges in the base network are randomly chosen and rewired to form irreciprocal edges. Specifically, in each step, a pair of reciprocal edges $i \leftrightarrow j$ is randomly chosen. If there is no link between two randomly chosen nodes $k$ and $l$ (\autoref{sup_fig2}(a)), then we randomly pick one of the two links between $i$ and $j$ and use it to connect $k$ and $l$ (\autoref{sup_fig2}(b)). Such a process leads to an indegree-outdegree correlation $\rho  \approx 1 - \lambda $. The rewiring process is restarted from the beginning if the network is disconnected.

\textbf{Step 3} \textit{Varying attractivity ratio}. The same process as described in \textbf{Step 3} in Section \ref{sectionNet1} is used to generate different $m^*$ values for each network generated in \textbf{Step 2}.

\begin{figure*}[ht]
\centerline{\includegraphics[width=.3\textwidth]{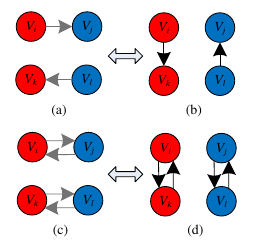}}
\caption{Net2: Illustration of the rewire process resulting in a change of $h_A$. Red: trait $A$, Blue: trait $B$}\label{sup_fig3}
\end{figure*}

\textbf{Step 4} \textit{Homophily}. In order to generate networks with different homophily for group $A$, $h_A$, links are further rewired in each network generated in \textbf{Step 3}, Let ${h'_A}$ be the homophily of group $A$ in the current network. At each step, either a pair of irreciprocal links or reciprocal links are randomly picked (\autoref{sup_fig3}); if ${h'_A} > h_A$, meaning that there are too many within-group connections, we rewire the within group links, $i \to k$, $l \to j$ (or $i \leftrightarrow k$, $l \leftrightarrow j$), to $i \to j$, $k \to l$  (or $i \leftrightarrow j$, $k \leftrightarrow l$), or vice versa if ${h'_A} < h_A$. The above process is repeated until ${h'_A} = h_A$.

\subsection*{A.3. Net3}
Net3 is the set of networks with different levels of indegree correlation ($\gamma \in [0, 0.4]$), varied from the MSM network; all four node properties (\textit{age}, \textit{county}, \textit{civil status} and \textit{profession}) are kept.

\textbf{Step1} \textit{Base network}. The MSM friendship network obtained from the web community~\cite{Rybski2009, Xin2011}.

\begin{figure*}[ht]
\centerline{\includegraphics[width=.3\textwidth]{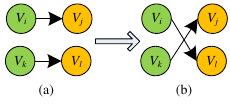}}
\caption{Net3: Illustration of the shuffling process used for network generation. $V_j$ and $V_l$ have the same property.}\label{sup_fig4}
\end{figure*}

\textbf{Step2} \textit{Varying $\gamma$}. A shuffling method slightly different from what was described in \cite{Xulvi2004} is used to generate networks with different indegree correlation. At each step, we randomly pick a pair of edges, $i \to j$, $k \to l$ (\autoref{sup_fig4}). If the indegrees of $i$ and $l$ are the two largest or the two smallest among the four nodes, and $j$ and $l$ have the same property, we rewire the two edges as $i \to l$, $k \to j$. Then, the degree distribution and homophily of the network is kept, and the indegree correlation increases as the rewiring process progresses. We generate networks with $\gamma$ up to 0.4 for each of the four properties in the MSM network.

\section*{Appendix B: Discussions on the estimate of $S_{XY}$ in $SH_{in}$}
It can be proven that when the network is undirected, a node will be recruited into a RDS sample with a probability proportional to its degree if the assumptions for $SH_{out}$ are fulfilled~\cite{Salganik2004,Volz2008}: $\{{\pi _i} \sim {d_i}/\sum\nolimits_i {{d_i}}\}$. Consequently, each edge in the network, $\{e_{i \to j}\}$, has a probability $\{{\pi _{i \to j}} = {\pi _i}/{d_i} \sim 1/\sum\nolimits_i {{d_i}}\}$ to be sampled, and the observed recruitment matrix from the RDS sample is an unbiased estimate of $S$.

However, when the network is directed, the inclusion probability for a node is no longer proportional to its degree, and the observed recruitment matrix from the sample will be representative only if individuals of the same group have similar edge formations, i.e., the personal recruitment matrix, $\{S_{XY}^i\}$, is the same for individuals in group $X$. Then, the observed raw recruitment matrix could be an appropriate estimate of ${S_{XY}}$.

A more general way is to develop a Hansen-Hurwitz type estimator for $S_{XY}$ using the edges' inclusion probabilities ($\{{\pi _{i \to j}} = {\pi _i}/{d_i^{out}}\}$):

\begin{equation}{\hat s_{XY}} = \frac{{\sum\nolimits_{i \to j,i \in X,j \in Y} {\frac{{d_j^{out}}}{{\pi _i}}} }}{{\sum\nolimits_{i \to j,i \in X} {\frac{{d_j^{out}}}{{\pi _i}}} }},\end{equation}

where $X$ and $Y$ are the set of nodes with corresponding properties in the sample. Since $\{\pi _i\}$ is usually not known when knowledge about the structure of the network is incomplete, we might use the mean field approach to approximate $\{\pi _i\}$ with the average inclusion probability for nodes within group $X$:

\begin{equation}
{\hat s_{XY}} = \frac{{\sum\nolimits_{i \to j,i \in X,j \in Y} {\frac{{d_j^{out}}}{{{\pi _X}}}} }}{{\sum\nolimits_{i \to j,i \in X} {\frac{{d_j^{out}}}{{{\pi _X}}}} }} = \frac{{\sum\nolimits_{i \to j,i \in X,j \in Y} {d_j^{out}} }}{{\sum\nolimits_{i \to j,i \in X} {d_j^{out}} }}.\label{eq_adj_S}
\end{equation}

Frankly, both using the observed recruitment matrix from the sample, and approximating as described by (\ref{eq_adj_S}), are brutal methods for estimating $S_{XY}$. We have tried both in the $SH_{in}$ estimator; however, it turns out that the adjustment for $\hat s$ made by (\ref{eq_adj_S}) always generates larger error and bias; we thus only provide the discussions here and choose not to show any results in the paper.

\section*{Appendix C: Supporting figures}

 \begin{figure*}[ht]
 \centerline{\includegraphics[width=.8\textwidth]{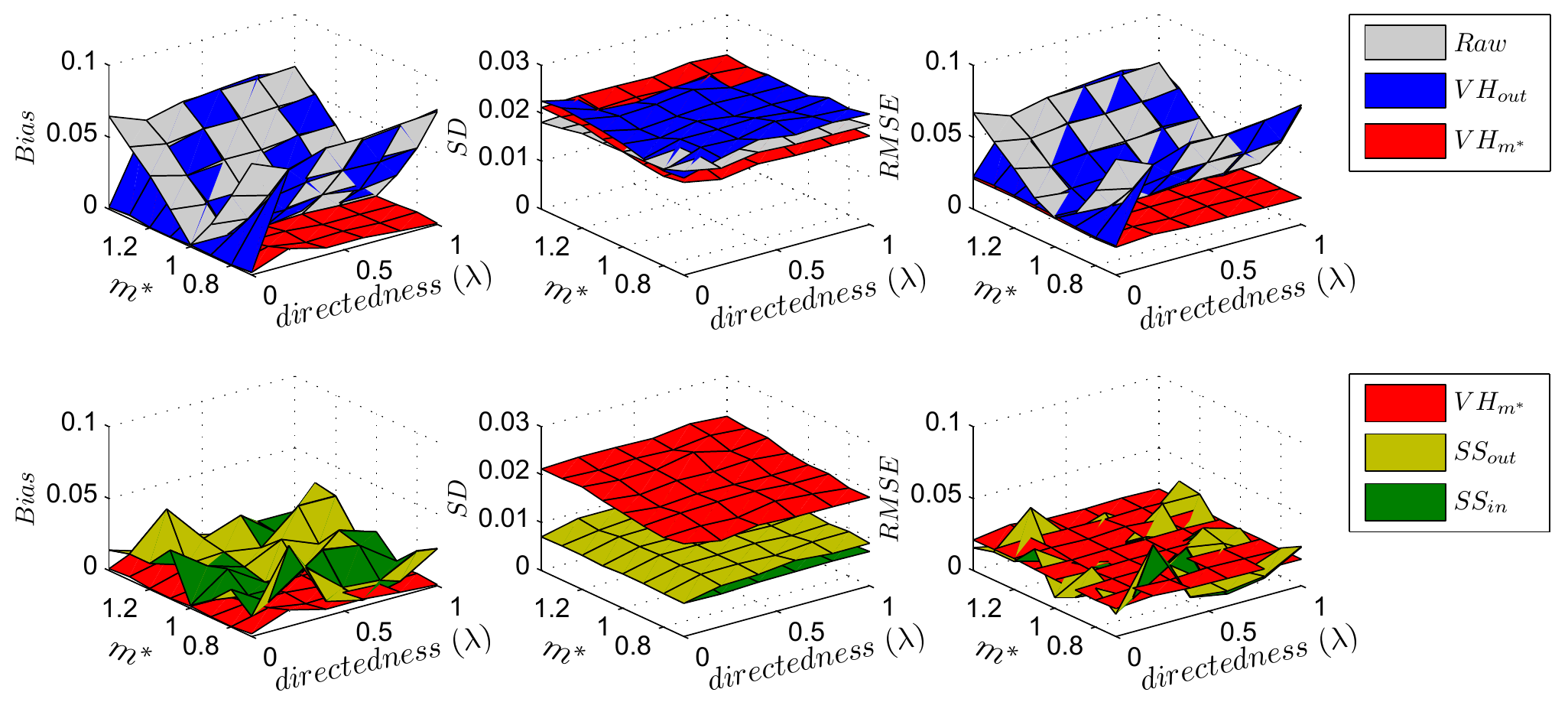}}
 \caption{Bias, Standard Deviation, and Root Mean Square Error of RDS estimators on Net1. Sampling without replacement, number of seeds=6, coupons=2, sample size=500.}
 \end{figure*}

 \begin{figure*}[ht]
 \centerline{\includegraphics[width=.8\textwidth]{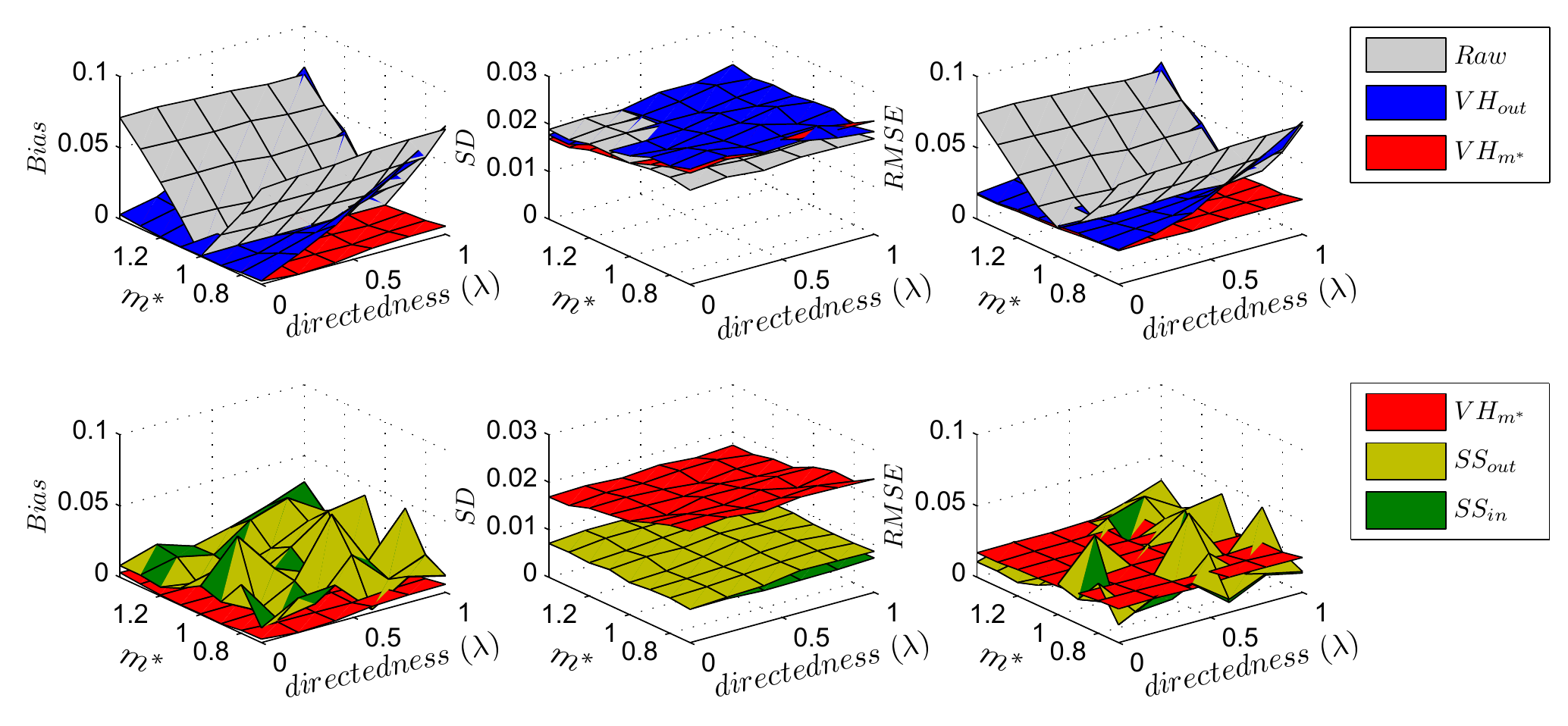}}
 \caption{Bias, Standard Deviation, and Root Mean Square Error of RDS estimators on Net2, homophily ${h_A} = 0$. Sampling without replacement, number of seeds=6, coupons=2, sample size=500.}
 \end{figure*}

 \begin{figure*}[ht]
 \centerline{\includegraphics[width=0.8\textwidth]{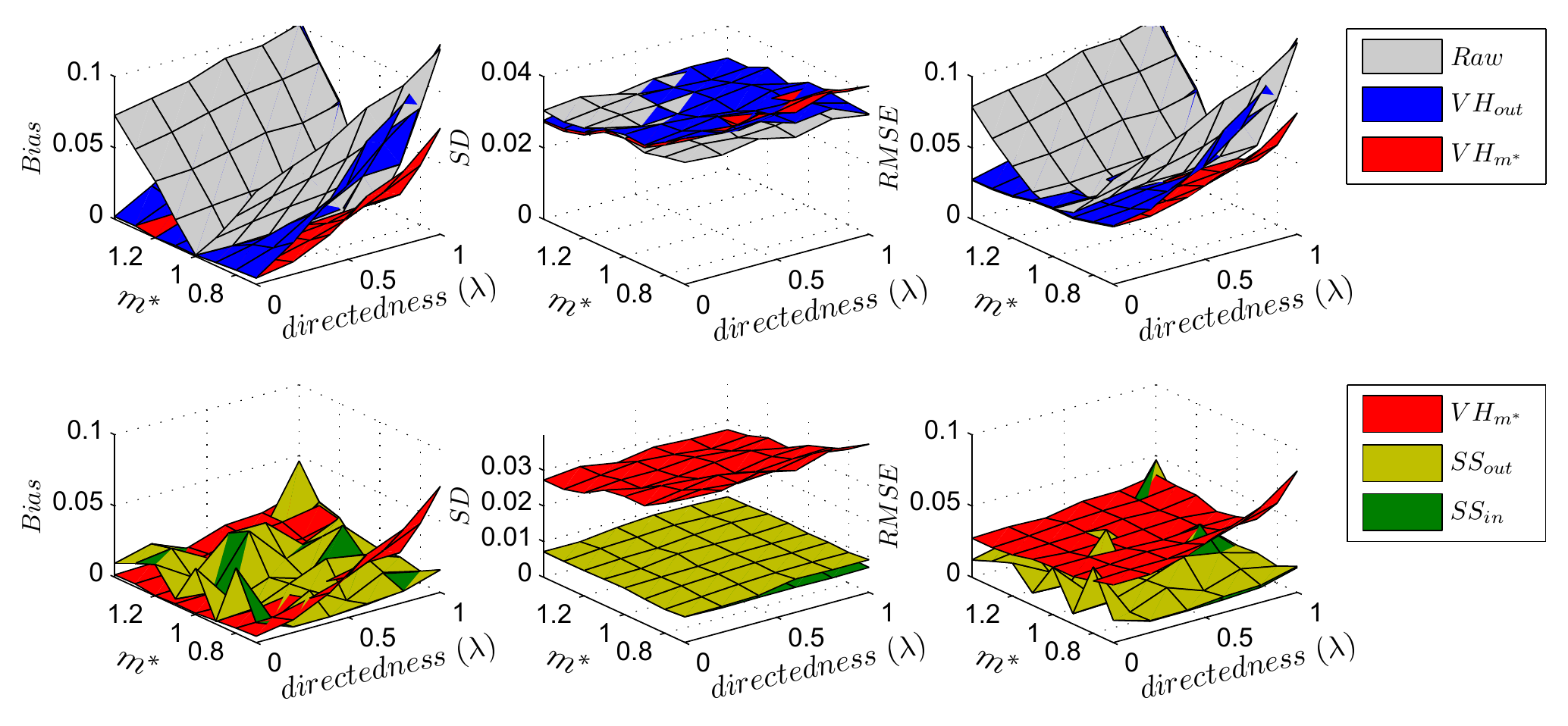}}
 \caption{Bias, Standard Deviation, and Root Mean Square Error of RDS estimators on Net2, homophily ${h_A} = 0.4$. Sampling without replacement, number of seeds=6, coupons=2, sample size=500.}
 \end{figure*}

 \begin{figure*}[ht]
 \begin{center}
 \centerline{\includegraphics[width=0.8\textwidth]{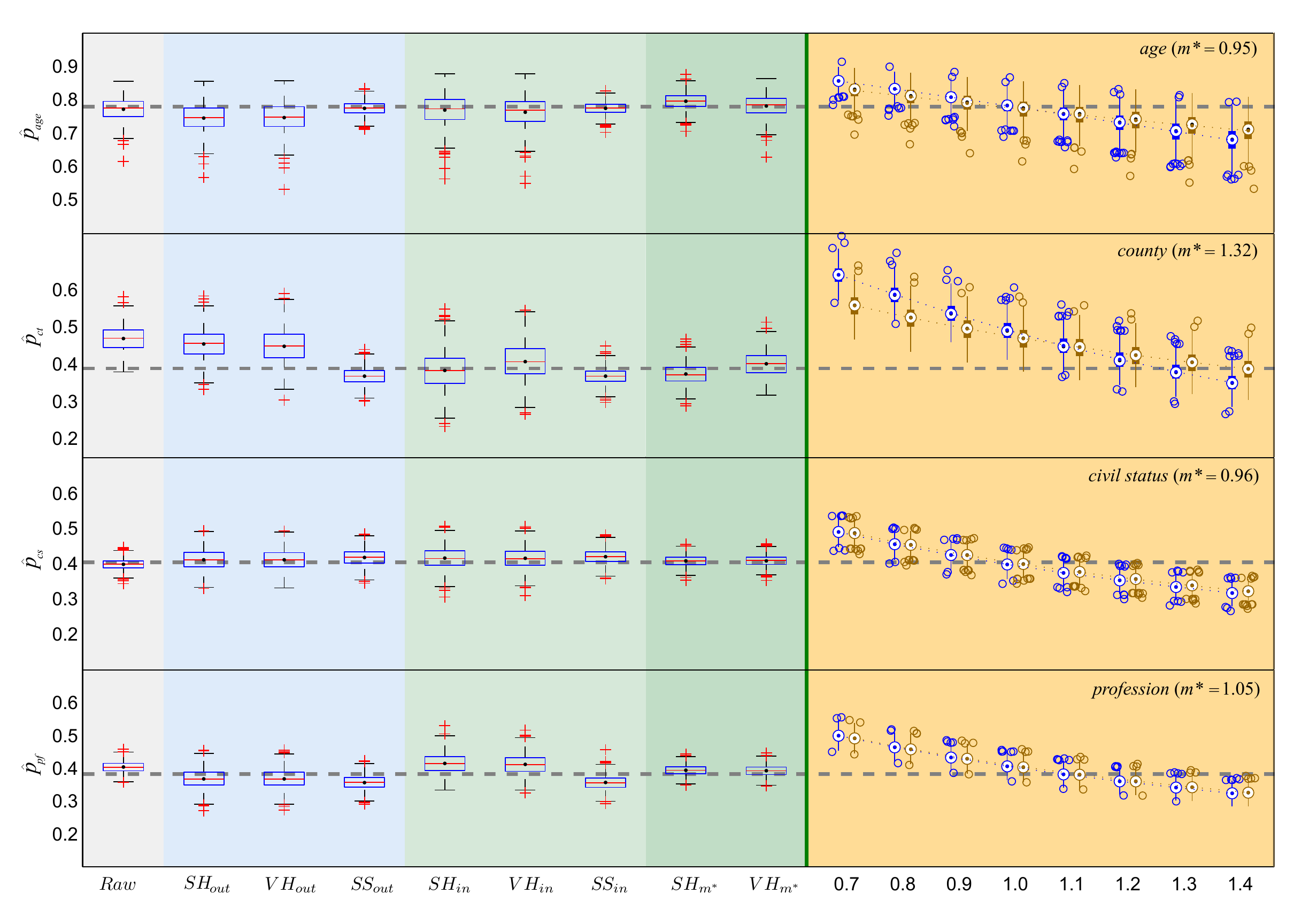}}
 \caption{RDS on MSM network. The right panel shows sensitivity analysis of $\hat p_{A}^{VH_m}$ (brown) and $\hat p_{A}^{SH_m}$  (blue) with $m$ varying from 0.7 to 1.4, plots are horizontally shifted a few points to avoid overlapping. Sampling with replacement, number of seeds=6, number of coupons=2, sample size=1000.}
 \end{center}
 \end{figure*}

 \begin{figure*}[ht]
 \begin{center}
 \centerline{\includegraphics[width=0.8\textwidth]{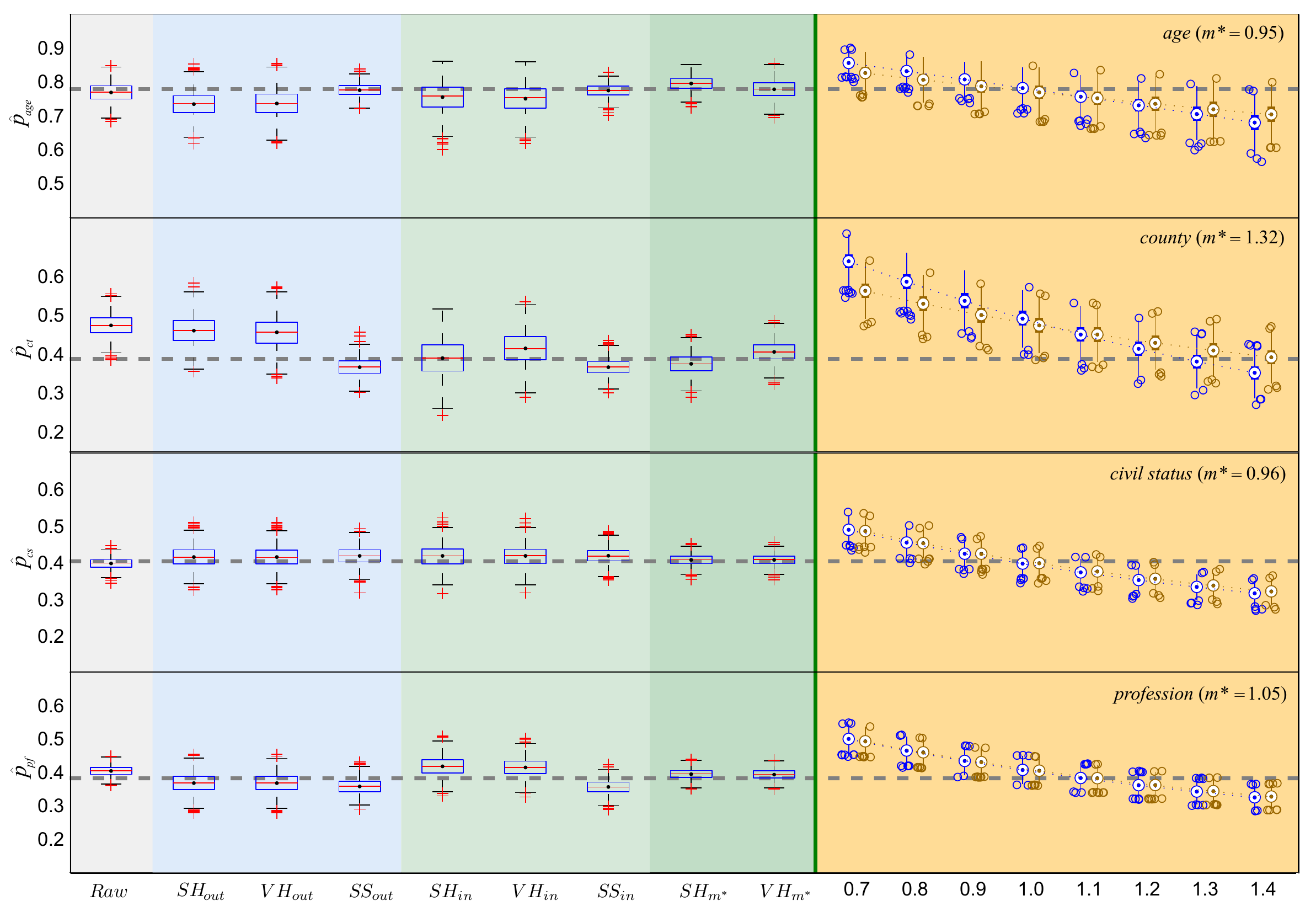}}
 \caption{RDS on Net3 with indegree correlation $\gamma  = 0.4$. The right panel shows sensitivity analysis of $\hat p_{A}^{VH_m}$ (brown) and $\hat p_{A}^{SH_m}$ (blue) with $m$ varying from 0.7 to 1.4, plots are horizontally shifted a few points to avoid overlapping. Sampling without replacement, number of seeds=6, number of coupons=2, sample size=1000.}
 \end{center}
 \end{figure*}

\clearpage

 \begin{figure*}[t]
 \begin{center}
 \centerline{\includegraphics[width=1\textwidth]{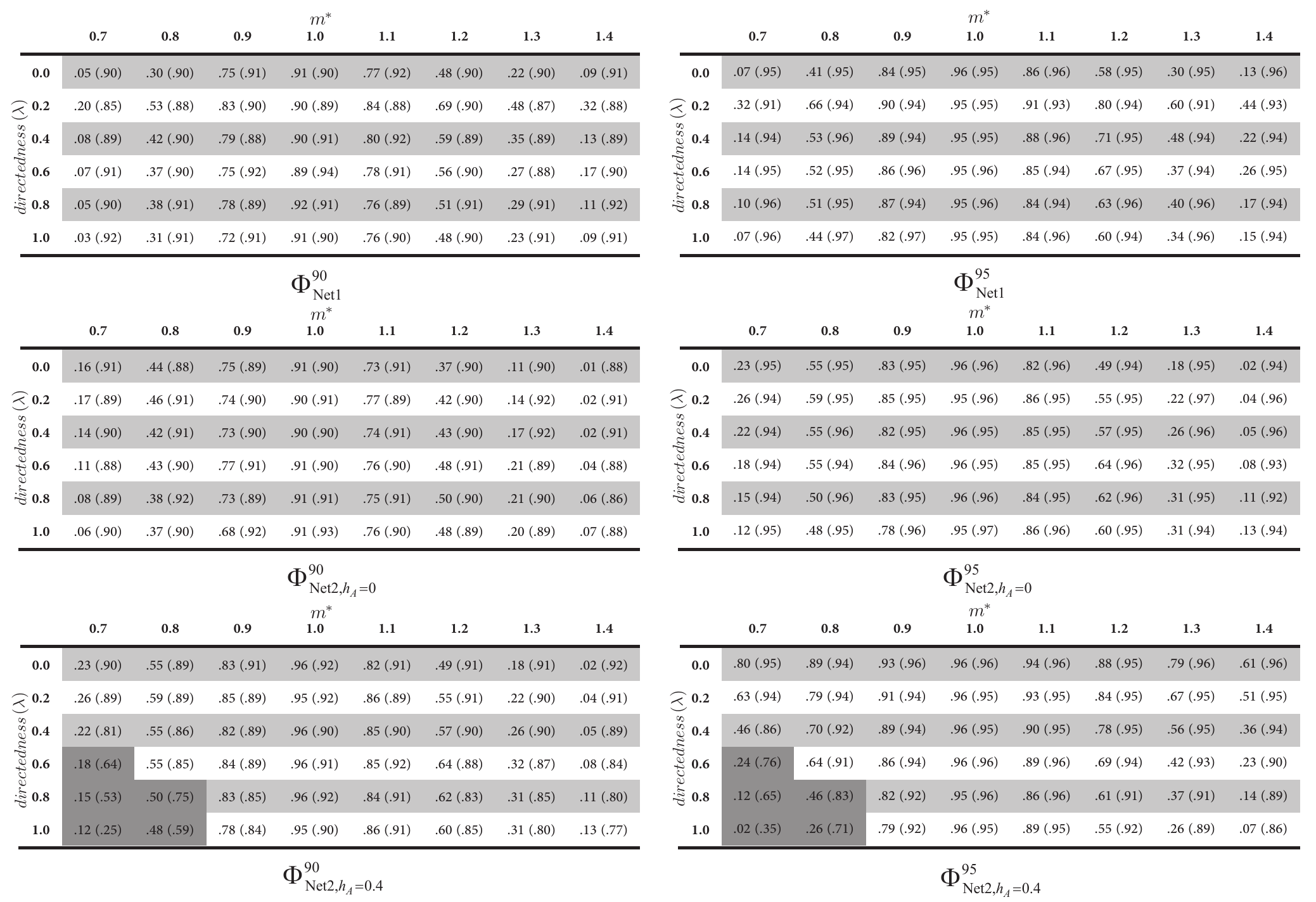}}
 \caption{90\% and 95\% bootstrap coverage probability of $\hat p_{A}^{VH_{out}}$ and $\hat p_{A}^{VH_{m^*}}$ (shown in brackets) on Net1 and Net2. Sampling without replacement, number of seeds=6, coupons=2, sample size=500.}
 \end{center}
 \end{figure*}

\bibliographystyle{elsarticle-num}
\bibliography{RDS_EJS}

\end{document}